\documentclass{aa} 
\usepackage{txfonts} 
\usepackage{natbib}
\usepackage{graphicx}
\usepackage{lscape} 
\usepackage{longtable}
\usepackage{array}

\begin{document}

\title{An X-ray spectral survey of the disc of M31 with XMM-Newton}
\subtitle{} 
\author{L. Shaw Greening\inst{1} \and R. Barnard\inst{1}
  \and U. Kolb\inst{1} \and C. Tonkin\inst{1} \and
  J.P. Osborne\inst{2}} 
\institute{The Department of Physics and
  Astronomy, The Open University, Walton Hall, Milton Keynes, MK7 6AA,
  UK \and The Department of Physics and Astronomy, The University of
  Leicester, Leicester, LE1 7RH, UK} 
\date{Received date / Accepted date}

\abstract{}{We present the results of a complete spectral survey of
  the X-ray point sources detected in five XMM-Newton observations
  along the major axis of M31 but avoiding the central bulge, aimed at
  establishing the population characteristics of X-ray sources in this
  galaxy.}{We obtained background subtracted spectra and lightcurves
  for each of the 335 X-ray point sources detected across the five
  observations from 2002. We also correlate our source list with those
  of earlier X-ray surveys and radio, optical and infra-red
  catalogues.  Sources with more than 50 source counts are
  individually spectrally fit in order to create the most accurate
  luminosity functions of M31 to date.}{Based on the spectral fitting
  of these sources with a power law model, we observe a broad range of
  best fit photon index.  From this distribution of best fit index, we
  identify 16 strong high mass X-ray binary system candidates in M31.
  We show the first cumulative luminosity functions created using the
  best fit spectral model to each source with more than 50 source
  counts in the disc of M31. The cumulative luminosity functions show
  a distinct flattening in the X-ray luminosity L$_\mathrm{X}$
  interval $37.0 \lesssim \log \mathrm{L_X}$ erg s$^{-1} \lesssim
  37.5$.  Such a feature may also be present in the X-ray populations
  of several other galaxies, but at a much lower statistical
  significance. We investigate the number of AGN present in our source
  list and find that above L$_\mathrm{X}\sim1.4\times10^{36}$ erg
  s$^{-1}$ the observed population is statistically dominated by the
  point source population of M31.}{}  \maketitle


\section{Introduction}
\label{sec:intro}
The Andromeda Galaxy (M31) is the nearest spiral galaxy to our own,
lying at a distance of 760 kpc \citep{vdBergh00}. The sources in M31
are observed at a nearly uniform distance and through an absorption
column significantly lower than for sources in the Galactic plane.
Thus M31 is an ideal target for studying the emission from the X-ray
point sources in a galaxy similar to the Milky Way.

M31 has been observed with many X-ray observatories since
\emph{Einstein}, when \citet{VS79} observed 69 point sources above
5$\times10^{36}$ erg\ s$^{-1}$. Two ROSAT surveys \citep{Supper97,
  Supper01} covered most of the M31 disc and found 560 sources above
5$\times10^{35}$ erg\ s$^{-1}$.  There have also been many
\emph{Chandra} \citep[e.g.][]{Williams04, Kong02, Kong03} and
XMM-Newton \citep[e.g.][]{Trud06, PFH05, Osborne01, Shirey01} surveys
of both the disc and central region of M31.

The X-ray emission from M31 is dominated by point sources mostly
consisting of X-ray binary systems (XBs). \citet{Trud06} surveyed 123
sources in the central region of M31 and reported that the majority
have X-ray properties reminiscent of Galactic low mass XBs (LMXBs),
and labelled 44 sources as XB candidates based on their spectral
properties and variability.

Six neighbouring, slightly overlapping XMM-Newton observations along
the major axis of M31 were made in January and June 2002.  These
observations, along with others taken between 2000 and 2007, form part
of a survey of the whole optical D$_{25}$ ellipse of M31. Since the
central region of M31 is well studied and the wider survey has not
been completed at the time of this work, we have investigated the five
remaining major axis observations that exclude the core region.
Henceforth we refer to these fields as the M31 disc fields. These
observations were long and uninterrupted; together with the
unprecedented effective area of XMM-Newton, they yielded up to 40
times the photon counts of the best previous observations. Previous
work on these XMM-Newton fields \citep[see e.g.][]{PFH05, Trud02} has
derived only spectral properties for the brightest few sources.

In this paper we re-analyse the five M31 disc fields. For the first
time we extend the spectral analysis to sources down to L$_\mathrm{X}
\gtrsim 10^{36}$ erg s$^{-1}$. We create a new source list, derive the
spectral parameters of each source and create spatially resolved
cumulative X-ray luminosity functions (CLFs). In Sect.~\ref{sec:obs}
we give details of the observations and data reduction,
Sect.~\ref{sec:analysis} covers the analysis and the results of cross
correlations with the \citet{PFH05} catalogue and catalogues at other
wavelengths.  In Sect.~\ref{sec:results} we give details of the
analysis of our spectral fitting including the creation of CLFs and
comments on the contamination of the CLF by background AGN.  Finally,
Sect.~\ref{sec:conclusion} summarises our findings.

\section{Observations and Data Reduction}
\label{sec:obs}

One observation of each disc field of M31 was taken using the EPIC pn
\citep{Struder01} and MOS \citep{Turner01} cameras on XMM-Newton in
January and June 2002; a journal of the observations is presented in
Table~\ref{tab:observations}. From north to south, we refer to the
fields as North 3, North 2, North 1, South 1 and South 2. Data were
processed using XMM-Newton SAS (version 6.5.0) tasks \texttt{epproc}
and \texttt{emproc} with up to date calibration. There are also
multiple observations of the central region taken between 2000 and
2005; these have been analysed by \citet{Trud02}, \citet{PFH05} and
\citet{Trud06} and are not covered here.

\begin{table*}
  \begin{center}
    \caption{Journal of XMM-Newton Observations of M31.  The
      field, observation number, date, pointing direction, filter
      used, total exposure (Exp) and exposure of the good time
      interval (GTI) and the number of sources detected are given.}
    \label{tab:observations}
    {\small
      \begin{tabular}{c c c c c c c c c}
        \noalign{\smallskip}
        \hline
        \hline
        \noalign{\smallskip}
        Field & Observation & Date & \multicolumn{2}{c}{Pointing
          direction} & Filter & Exp & GTI & Sources \\
        & & & \multicolumn{2}{c}{RA/dec (J2000)} & & ks & ks & \\
        \noalign{\smallskip}
        \hline
        \noalign{\smallskip}
        North 3 (n3) & 0109270401 & 29 June 2002 & 0:46:38 & +42:16:20 & Medium & 55 & 46 & 80 \\
        North 2 (n2) & 0109270301 & 26 Jan 2002 & 0:45:20 & +41:56:09 & Medium & 55 & 25 & 57 \\
        North 1 (n1) & 0109270701 & 05 Jan 2002 & 0:44:01 & +41:35:57 & Medium & 55 & 55 & 82 \\
        South 1 (s1) & 0112570201 & 12 Jan 2002 & 0:41:25 & +40:55:35 & Thin & 53 & 44 & 71 \\
        South 2 (s2) & 0112570301 & 24 Jan 2002 & 0:40:06 & +40:35:24 & Thin & 58 & 24 & 45 \\
        \noalign{\smallskip}
        \hline
        \noalign{\smallskip}
      \end{tabular}
      }
  \end{center}
\end{table*}

\subsection{Source Detection}
\label{ssec:detection}
For the purposes of source detection, the observations were screened
for periods of high background counts in each camera. Lightcurves
including all counts above 10 keV were created for each camera, and
intervals with levels above 1 count s$^{-1}$ for the pn and 0.5 counts
s$^{-1}$ for each of the MOS cameras were excluded. Observations were
then synchronised and source detection carried out. For the source
detection the data were split into 5 energy bands: (0.2-0.5) keV,
(0.5-2) keV, (2-4.5) keV, (4.5-7) keV and (7-12) keV.  For the pn data
we used only ``single'' events (PATTERN==0) in the first energy band
and for the other bands ``singles and doubles'' were selected
(PATTERN$ <=4$). Additionally, for the pn, only events with (RAWY$
>12$) for the pn were used and to avoid emission from the spatially
inhomogeneous Copper fluorescent line, the energy range (7.8-8.2) keV
was omitted from band 5. For MOS data ``singles'' to ``quadruples''
(PATTERN$ <=12$) were selected. For each camera, source lists were
constructed in each energy band using
\texttt{edetect\_chain}\footnote{http://xmm.vilspa.esa.es/sas/6.5.0/doc/edetect\_chain/index.html}
with a minimum likelihood threshold of 10. These lists were then
combined to form a final source list.  All source regions were
  then set to have a radius of 40$''$ as this corresponds to $\sim
  88\%$ encircled energy at 1.5~keV. Finally the sources in this list
were visually inspected for overlapping source regions. When a 40$''$
source extraction region contained more than one source, the region
was reduced to 20$''$. Any 20$''$ extraction region containing
  more than one source was deleted.

\subsection{Background Selection}
\label{ssec:background}
Backgrounds were selected for each source based on the following
criteria. Suitable backgrounds must be on the same CCD as the source,
have no sources within the background region, and must have a lower
count density (fewer counts per unit area) than the source region. The
latter criterion ensures that there are no unresolved faint sources or
areas of diffuse emission in the background region that combine to an
anomalously high count density. For source regions on a chip gap or
chip edge background regions must be on the same chip edge or gap and
have the same percentage of off-chip area as the source region.
Finally, background regions have a radius between one and four times
the radius of the source extraction region.

\subsection{Scientific Product Extraction}
\label{ssec:products}
Following \citet{Barnard07var}, synchronised source and background
lightcurves with 2.6~s time resolution were extracted from each of the
three detectors.  These were summed to give a combined, background
subtracted EPIC lightcurve for every source.  

Energy spectra were extracted from the source and background regions
with 5~eV binning for the pn camera and with 15~eV binning for the MOS
cameras.  A response matrix (RMF) and ancillary response file (ARF)
were also generated for each source spectrum. For any source with
spectra from both the MOS cameras we added together the two spectra to
form a combined MOS spectrum, otherwise just the one MOS spectrum was
used in the following analysis. Counts outside the 0.3-10 keV range
were rejected.

\section{Analysis}
\label{sec:analysis}
There were 335 point source detections with a minimum likelihood of 10
in the 5 disc fields of M31. These sources are ordered by RA and we
present their positions and X-ray properties in
Table~\ref{tab:sources}. Of these 335 detections, 6 sources were
detected in two observations and so there are 329 distinct point
sources across the disc. Their count rates range from $3.42 \times
10^{-5}$ to 0.403 counts per second.

\subsection{Lightcurves}
Lightcurves were binned to 100, 200 and 400 second bins and checked
for variability by examining how well they were fit by a line of
constant intensity. Sources with a null hypothesis probability of
$>5~\%$ in any of the lightcurves were classed as variable. This is a
conservative approach that may miss variability in the data, but
reduces the number of sources with spurious variability.  Around 300
of the M31 disc sources were too faint to detect variability on the
timescales sampled. The lightcurves were also visually inspected for
bursts, dips or other behaviour. Only source 238 (identified as a
flare star by Pietsch et al. 2005 (their source 663) and by
Trudolyubov et al. 2005 (their source 22) was found to be variable,
with $30 \pm 3 \%$ rms variability.

\subsection{Energy Spectra}
\label{sec:models}
222 of the sources in the disc of M31 have spectra with sufficient
photons ($>$50 source counts in the pn or combined MOS) to allow
spectral fitting, the results of which are given in
Table~\ref{tab:sources}.

We binned the pn and MOS spectra depending on source intensity.
Spectra exceeding 500 source counts over the observation were grouped
to a minimum of 50 counts per bin.  Spectra containing between 200 and
499 source counts were grouped to a minimum of 20 counts per bin.
Spectra with between 50 and 199 source counts and with more than 50\%
of the total counts from the source were grouped to a minimum of 10
counts per bin, while those with between 50 and 199 source counts but
with less than 50\% of the total counts from the source were also
grouped to a minimum of 20 counts per bin. Each grouped energy
spectrum was freely fit by three spectral models: blackbody,
bremsstrahlung and power law emission models, using xspec
11.3.1\footnote{http://heasarc.gsfc.nasa.gov/docs/xanadu/xspec/index.html}.
Sources which have very few or no counts above 2~keV were also fit
with a neutron star atmosphere (nsa) model which resembles the
emission from a super soft source.  For all the models the absorption
was a free parameter but with a minimum of at least $0.1 \times
10^{22}$ cm$^{-2}$, the Galactic foreground absorption
\citep{Dickey90}. The source flux was calculated from the best fit
model.  The spectral parameters of each source give a greater insight
into its properties than its X-ray hardness ratios or variability
alone.  Sources with less than 50 source counts in both cameras are
dealt with on a field by field basis as described below.

\subsection{Faint Sources}
\label{ssec:faint}

\begin{table*}
  \begin{center}
    \caption{Best fit parameters for power law models applied to the
      summed spectra of the faint sources in each of the disc fields.
      We show the total number of faint sources in each field, the
      power law photon index, $\Gamma$, $\chi^2/$dof and for which
      camera this fit was found. The absorption was fixed to $0.10
      \times10^{22}$ H atom cm$^{-2}$ for each field.  The bracketed
      numbers are the error in the last significant figure.  Where we
      quote two conversion factors, the first is for the pn camera and
      the second for the MOS. Errors are unavailable for the best fit
      power law index to South 1 as the $\chi^2/$dof is $>2$.}
    \label{tab:pimms}
    \begin{tabular}{c c c c c c}
      \noalign{\smallskip}
      \hline
      \hline
      \noalign{\smallskip}
      Field & Number of & $\Gamma$ & $\chi^2/$dof & Camera(s) used & Conversion Factor\\
       & faint sources & & & & erg s$^{-1}$/counts s$^{-1}$ \\
      \noalign{\smallskip}
      \hline
      \noalign{\smallskip}
      North 3 & 19 & 2.8(13) & 161/122 & pn / MOS & $5.71\times10^{38}$ / $9.33\times10^{38}$\\
      North 2 & 32 & 1.5(4) & 39/38 & pn / MOS & $3.58\times10^{38}$ / $7.82\times10^{38}$\\
      North 1 & 11 & 1.00(15) & 24/16 & MOS & $5.73\times10^{38}$\\
      South 1 & 16 & 1.3(-) & 450/120 & pn & $4.75\times10^{38}$\\
      South 2 & 17 & 1.3(4) & 56/46 & pn & $4.54\times10^{38}$\\
      \noalign{\smallskip}
      \hline
      \noalign{\smallskip}
      \end{tabular}
    \end{center}
\end{table*}

For the 95 detections with too few photons to allow spectral fitting
($<$ 50 source counts in both cameras), the parameters of the best fit
absorbed power law for the field were used. First we grouped the
spectra according to field and camera (pn or MOS), creating 10 groups.
The spectra of the faint sources in each group were then summed to
give one spectrum for each camera's observation of every field. The
absorption was fixed to 0.1 $\times 10 ^{22}$ H atoms cm$^{-2}$ and
the best fit photon index to the summed spectrum was used to calculate
a count rate to flux conversion for that camera's observation of the
field. For South 1 we do not quote errors on the photon index because
the $\chi^2/$dof is $>2$. Although this is not a good fit to the South
1 sources, it is the best fit power law and so we have used it to
maintain consistency across the fields.

Some sources are detected only on the pn or one of the MOS
cameras and not in all three. For some fields only one of the two
summed spectra (pn or MOS) could be well fit using a power law model.
Where a faint source is detected in one camera but a good fit to the
summed faint source spectrum for that field is only available for the
other camera we do not give a source luminosity.

The parameters of the best fit power law to each field are given in
Table~\ref{tab:pimms}. For the faint sources the quoted conversion
factor was applied to the exposure corrected count rate. There is an
obvious change in the photon index in the northern disc: with
increasing distance from the centre of M31 the best fit power law
becomes softer.  However we caution against drawing conclusions from
this as it is based on the summed spectrum of a small number of faint
sources and the photon indicies are consistent with each other within
errors.

We have calculated the 0.3-10~keV luminosity from either a source
spectrum or from an average model for the relevant field, for 317 of
the 329 sources, and these are given in Table~\ref{tab:sources}.

\subsection{Cross-correlations with other M31 catalogues}
\label{subsec:counterparts}
We searched for cross-correlations within a radius of
$3({(\sigma_{statistical})^2 + (\sigma_{systematic})^2})^{1/2}$,
where, for the uncorrected XMM-Newton positions from this survey,
$\sigma_{statistical}=1''$ and $\sigma_{systematic}=3''$. The
statistical error is taken from the 2XMM
catalogue\footnote{http://xmmssc-www.star.le.ac.uk/Catalogue/UserGuide\_xmmcat.html}.
This error is strongly dependent on source counts, however for our
range of source count rates we have assumed a representative value of
1~arcsecond for the statistical error.  The systematic error is
derived from the offset to each field. The most accurate XMM-Newton
positions have residual systematic errors of around $0.5''$, and it
can be seen from \citet{PFH05} that the M31 disc fields each have an
additional offset of $0.3-2.3''$.  Thus we have used a conservative
systematic error of $3''$. This gives a search radius of 10$''$.  For
295 out of our 329 sources we found a source within the search radius
in the \citet{PFH05} catalogue, and a summary of the classifications
of these sources as determined in \citet{PFH05} are given in
Table~\ref{tab:PFH05counter}. Sources are either classed as
``candidates'' or ``members of'' each class in \citet{PFH05}, but here
they are grouped together. The hard class contains all the sources
with HR2-EHR2 $> -0.2$ or only HR3 and/or HR4 defined, and no other
classification \citep[see][for the definitions of hardness ratios HR2,
HR3, HR4 and EHR2 and full details]{PFH05}.

For the 34 sources not in \citet{PFH05} we searched the following
catalogues for counterparts:

(i) \emph{X-ray sources:} the \emph{Einstein} \citep{TF91}, ROSAT/PSPC
\citep{Supper97, Supper01} and \emph{Chandra} \citep{Williams04,
  Kaaret02} catalogues.

(ii) \emph{Stellar objects:} USNO-B1 \citep{USNOB1}, 2MASS
\citep{2MASS} and the Local Group Survey \citep{LGSM31}.

(iii) \emph{Radio sources:} VLA all sky catalogue \citep{Condon98} and
catalogue of sources within M31 \citep{Walterbos85}.

(iv) \emph{Globular cluster candidates:} the Bologna catalogue
\citep{Galleti04} and the catalogue by \citet{Kodaira04}.

(v) \emph{Supernova remnant candidates:} catalogues by
\citet{Magnier95} and \citet{FJ78}.

Only 2 of the 34 sources in this survey are not found in any other
catalogue listed above. Eight sources are identified in either
\citet{Supper97} or \citet{Supper01}, one of which (source 184) is
identified as a variable supernova remnant (SNR). Two radio sources
were found in \citet{Condon98} (Sources 184 and 23), both of these are
also in the \citet{Supper97} or \citet{Supper01} source lists. All the
other sources have potential counterparts in the optical catalogues
and are classified as $<$hard$>$ following the convention of
\cite{PFH05}.

The classification for all 327 sources are given in
Table~\ref{tab:sources}; we distinguish the sources classified in this
work by a $^{(1)}$ beside their classification.

\begin{table}
  \begin{center}
    \caption{Summary of classifications of our sources from
      \citet{PFH05}.}
    \label{tab:PFH05counter}
    \begin{tabular}{c c}
      \noalign{\smallskip}
      \hline
      \hline
      \noalign{\smallskip}
      Type & Number \\
      \noalign{\smallskip}
      \hline
      \noalign{\smallskip}
      hard & 207 \\
      foreground star &  42 \\
      AGN/Galaxy & 19 \\
      Supernova remenants & 11 \\
      Globular cluster source & 13 \\
      Supersoft source & 2 \\
      X-ray binary system & 1 \\
      \noalign{\smallskip}
      \hline
      \noalign{\smallskip}
    \end{tabular}
  \end{center}
\end{table}

\section{Results}
\label{sec:results}
We present a summary of the results of our spectral analysis, with the
number of detections and faint sources per field as well as a
breakdown of the best fit models in each field, in
Table~\ref{tab:data}. The quoted luminosity of the faint limit in
Table~\ref{tab:data} is the luminosity of the brightest source with
less than 50 source counts. Full details of the source positions,
spectral fitting and classification (see
Sec.~\ref{subsec:counterparts}) of each source are listed in Table
\ref{tab:sources}.


\begin{table*}
  \begin{center}
    \caption{Overview of the source spectral analysis for each field.
      ``Faint'' denotes the number of sources with $<$50 source
      counts, in brackets is the number of sources for which we cannot
      quote a luminosity. For details on the HMXB candidates see text.
      The faint limit is the luminosity below which all sources have
      less than 50 source counts.}
    \label{tab:data}
    \begin{tabular}{c c c c c c c c c}
      \noalign{\smallskip}
      \hline
      \hline
      \noalign{\smallskip}
      Field & \multicolumn{7}{c}{Number of sources best fit by} & Luminosity of faint \\
      & detections & faint & power law & blackbody & bremsstrahlung
      & nsa &  HMXB candidates & limit, erg s$^{-1}$ \\
      \noalign{\smallskip}
      \hline
      \noalign{\smallskip}
      North 3 & 82 & 19 (3) & 55 & 2 & 5 & 0 & 7 & 8.6 $\times$10$^{35}$ \\
      North 2 & 57 & 32 (0) & 18 & 4 & 3 & 0 & 3 & 1.7 $\times$10$^{36}$ \\
      North 1 & 80 & 11 (2) & 55 & 8 & 6 & 0 & 3 & 5.0 $\times$10$^{35}$ \\
      South 1 & 71 & 16 (5) & 41 & 9 & 4 & 1 & 3 & 7.4 $\times$10$^{35}$ \\
      South 2 & 45 & 17 (3) & 23 & 4 & 0 & 1 & 2 & 9.8 $\times$10$^{35}$ \\
      \noalign{\smallskip}
      \hline
      \noalign{\smallskip}
      \end{tabular}
    \end{center}
\end{table*}

\subsection{Spectral Properties}
\label{subsec:pho_properties}
\begin{figure}
\begin{center}
\includegraphics[scale=0.3, angle=-90]{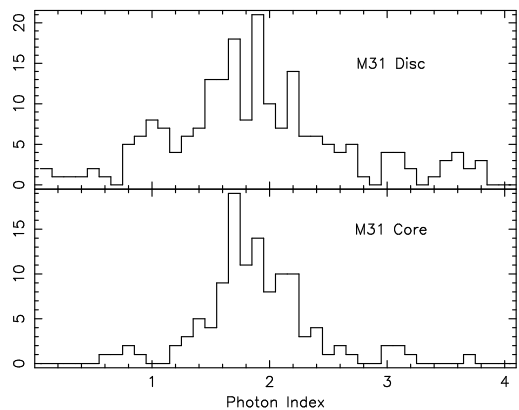}
\end{center}
\caption{Distribution of photon index derived from the best fit power
  law model for each source, with the exception of very soft sources
  ($\Gamma > 4$).  The M31 disc data is shown in the top panel and the
  data from the central region from \citet{Trud06} is shown in the
  lower panel. The bin width is 0.1.}
\label{fig:histo}
\end{figure}

Table \ref{tab:data} shows that a power law model is the best fit
model in the majority of cases, although Fig.~\ref{fig:histo} shows a
wide range of best fit photon index. Only four sources with more than
50 source counts have few or no counts above 2keV. These are fit with
the nsa model. Of these four only two sources are best fit with the
nsa model, accordingly these sources are described as supersoft.  Of
the three models tested in this survey (see Sec.~\ref{sec:models})
foreground sources would be best fit by a blackbody model; in fact 10
of the 27 sources best fit by a blackbody model in this work are
classified as foreground stars in \citet{PFH05} (see
Table~\ref{tab:sources}).

Following the work of \citet{Trud06} on the central region of M31, we
investigated the distribution of photon index for all sources, not
just those for which a power law was the best fit.
Figure~\ref{fig:histo} shows the distribution of the best fit power
law photon index ($\Gamma$) to every source. We have compared the data
from the disc to sources in the central region from \cite{Trud06}, who
used the same method to derive the spectral indices except that they
have presented a weighted mean of the spectral indices derived from
multiple observations. There are 33 disc sources with extremely soft
spectra ($\Gamma > 4$) which are not plotted.

Figure~\ref{fig:histo} shows a broad peak in both the central region
and the disc fields at $\Gamma\sim1.7$ which is expected for low mass
X-ray binary systems \citep[see][]{Trud06}.  However the disc sources
also show a peak at $\Gamma\sim1$, a feature which is absent in the
central region. While this excess is seen most obviously in the full
disc sample, it is also seen in each field individually. The excess is
suggestive but a KS test does not rule out that both samples in
Fig.~\ref{fig:histo} are drawn from the same parent population (KS
probability 0.36)

There are 23 disc sources with $0.8 \leq \Gamma \leq 1.2$ and seven of
these have very large errors to the best fit photon index.  The
variation in $\chi^2$ around the best fit photon index was
investigated for all 23 sources.  Five of the seven with large errors
were found to have a very shallow variation in $\chi^2$. This
indicates that the best fit $\Gamma$ is not well defined, we thus
exclude these sources from the group with $0.8 \leq \Gamma \leq 1.2$.
A photon index of around 1 is expected from magnetically accreting
neutron stars \citep{White05} and thus indicates the presence of
highly magnetic neutron stars in the disc of M31. Since the magnetic
field of the neutron star is expected to be weaker in LMXBs than HMXBs
\citep{Tauris06}, these sources are strong HMXB candidates - the first
such candidates in M31. The 18 HMXB candidates are given in Table
\ref{tab:HMXB}; we have 16 good candidates with small errors and two
secondary candidates with a large 90\% confidence interval but a
sharply defined minimum. Table~\ref{tab:HMXB} gives astrometrically
corrected positions from the source catalogue of \citet{PFH05} where
available.  For the three HMXB candidates not in the source catalogue
of \citet{PFH05}, we apply the appropriate astrometric corrections as
given by \citet{PFH05} to calculate the positions quoted in
Table~\ref{tab:HMXB}. For each of the HMXB candidates we also give the
details of the best fit power law model (n$_\mathrm{H}$ and $\Gamma$),
the luminosity derived from that fit and the V magnitude extinction,
calculated from the X-ray band absorption via A$_\mathrm{v} =
$n$_\textrm{H}$/($1.79\times10^{21}$) cm$^{-2}$, see
\citet{Predehl95}. From the V magnitude extinction it is also possible
to calculate the B$-$V colour excess E(B$-$V) = A$_\mathrm{v}$/3.24.

\begin{table*}
  \begin{center}
    \caption{X-ray properties of the 18 candidate HMXBs with a photon
      index between 0.8 and 1.2.  Coordinates are astrometrically
      corrected. The parameters of the best fit power law model are
      quoted in columns 5 \& 6. The symbol f signifies that the
      absorption was fixed to $0.1 \times10^{22}$ H atoms cm$^{-2}$
      for that field.  90\% confidence interval errors are quoted for
      both the absorption and photon index. Next we give the
      luminosity derived from the best fit model, the bracketed
      numbers are the error in the last significant figure. The
      luminosity and its errors are calculated from the 90\%
      confidence interval.  We also give the V magnitude absorption at
      the distance of M31.  The final column indicates the
      \citet{LGSM31} designation of any optical source within 3.3$''$
      of the astrometrically corrected position. The strong candidates
      are listed in the top section of the table and the secondary
      candidates in the lower section.}
    \label{tab:HMXB}
    \begin{tabular}{c c c c c c c c c}
      \noalign{\smallskip}
      \hline
      \hline
      \noalign{\smallskip}
      Field & Source & RA & Dec & n$_\mathrm{H}$ & Photon index & Luminosity & A$_\mathrm{v}$ & Optical \\
       & Number & (J2000) & (J2000) & $/10^{22}$ H atom cm$^{-2}$ & & $/10^{36}$ erg s$^{-1}$ & & coincidence \\
      \noalign{\smallskip}
      \hline
      \noalign{\smallskip}
      South 2 & 21 & 0:40:01.50 & +40:32:45.9 & $0.4^{+0.6}_{-0.3}$ & 0.9$^{+0.4}_{-0.5}$ & 2.3(10) & 2.29 & - \\
      \noalign{\smallskip}
      South 2 & 34 & 0:40:17.07 & +40:48:40.7 & $0.21^{+0.67}_{-0.14}$ & 1.2$^{+1.0}_{-0.7}$ & 8(4) & 1.17 & - \\
      \noalign{\smallskip}
      South 1 & 99 & 0:42:10.97 & +41:06:47.6 & f & 0.8$^{+0.5}_{-0.4}$ & 9(4) & 0.56 & J004210.83+410647.2 \\
      \noalign{\smallskip}
      South 1 & 106 & 0:42:16.76 & +41:00:21.0 & $0.4^{+0.6}_{-0.3}$ & 1.1$^{+0.6}_{-0.4}$ & 5(2) & 2.12 & - \\
      \noalign{\smallskip}
      North 1 & 123 & 0:43:01.44 & +41:30:17.5 & $0.18^{+0.07}_{-0.06}$ & 0.9$^{+0.1}_{-0.1}$ & 74(6) & 1.01 & J004301.51+413017.5 \\
      \noalign{\smallskip}
      North 1 & 149 & 0:43:54.50 & +41:31:04.0 & f & 0.9$^{+0.7}_{-0.7}$ & 1.0(7) & 0.56 & J004354.62+413101.0 \\
      \noalign{\smallskip}
      North 1 & 160 & 0:44:06.64 & +41:38:57.8 & f & 1.0$^{+0.6}_{-0.7}$ & 1.1(8) & 0.56 & - \\
      \noalign{\smallskip}
      North 1 & 172 & 0:44:20.87 & +41:35:41.9 & f & 1.1$^{+0.6}_{-0.6}$ & 1.3(9) & 0.56 & J004421.01+413544.3 \\
      \noalign{\smallskip}
      North 1 & 197 & 0:44:47.29 & +41:44:12.6 & f & 1.0$^{+0.8}_{-0.7}$ & 5(3) & 0.56 & - \\
      \noalign{\smallskip}
      North 2 & 256 & 0:45:58.82 & +42:04:27.5 & $0.3^{+0.4}_{-0.2}$ & 1.2$^{+0.4}_{-0.5}$ & 5(2) & 1.62 & J004558.98+420426.5 \\
      \noalign{\smallskip}
      North 3 & 234 & 0:45:34.96 & +42:17:53.0 & f & 0.9$^{+0.7}_{-0.6}$ & 4(3) & 0.56 & J004534.90+421752.8 \\
      \noalign{\smallskip}
      North 3 & 236 & 0:45:37.31 & +42:12:33.4 & f & 0.9$^{+0.6}_{-0.5}$ & 3(2) & 0.56 & - \\
      \noalign{\smallskip}
      North 3 & 294 & 0:46:43.80 & +42:09:48.2 & f & 1.1$^{+0.4}_{-0.3}$ & 1.9(10) & 0.56 & J004644.02+420950.0 \\
      \noalign{\smallskip}
      North 3 & 295 & 0:46:42.82 & +42:27:16.3 & f & 1.1$^{+0.7}_{-0.5}$ & 3(2) & 0.56 & - \\
      \noalign{\smallskip}
      North 3 & 302 & 0:46:53.52 & +42:19:14.4 & f & 0.9$^{+0.6}_{-0.7}$ & 1.1(7) & 0.56 & J004653.49+421914.4 \\
      \noalign{\smallskip}
      North 3 & 305 & 0:46:58.61 & +42:24:15.5 & f & 1.1$^{+0.8}_{-0.7}$ & 2.0(14) & 0.56 & - \\
      \noalign{\smallskip}
      \hline
      \noalign{\smallskip}
      South 2 & 9 & 0:39:38.77 & +40:47:55.9 & f & 0.8$^{+0.9}_{-0.7}$ & 5(4) & 0.56 & - \\
      \noalign{\smallskip}
      North 2 & 226 & 0:45:26.66 & +41:56:35.3 & f & 0.9$^{+0.9}_{-0.9}$ & 1.7(13) & 0.56 & J004526.58+415633.1 \\
      \noalign{\smallskip}
      \hline
      \noalign{\smallskip}
      \end{tabular}
    \end{center}
\end{table*}

The most recent complete optical survey of M31 has been the UBVRI
Local Group Survey of \citet{LGSM31}.  This survey consists of UBVRI
and various narrow band coverage of the entire optical D$_{25}$
ellipse of M31 down to a limiting magnitude of V=24.9. To search for
optical counterparts to our 18 HMXB candidates we adapted the criteria
for Galactic luminous Be stars from \citet{Sabogal05}. Galactic Be
stars have V-band absolute magnitudes between M$_\mathrm{V}$ -6 and 0,
as well as both $-0.4<$ B-V $<0.8$ and $-0.35<$ V-I $<0.8$ colour
restrictions. After correcting these criteria for the distance to M31
(V magnitude $>$ 18 and (B$-$V) $<$ 0.8) we searched the catalogue of
\citet{LGSM31} for possible counterparts within 3.3$''$ of the
astrometrically corrected positions.  This search radius is calculated
as in Sec.~\ref{subsec:counterparts} where, for the astrometrically
corrected XMM-Newton positions, $\sigma_{satistical}=1''$ and
$\sigma_{systematic}=0.5''$. Eight of the 16 good HMXB candidates and
one of the two secondary candidates have counterparts within this
search radii in \citet{LGSM31}, four of these have the magnitudes and
colours that we would expect for a Be-type star in M31. All
  potential counterparts are listed in Table~\ref{tab:HMXB}.

Using the method described below we investigated the possible
contamination of this potential HMXB population by background AGN.  We
find that AGN could make up $\sim 60\%$ of the total disc population
with L$_\mathrm{X} > 10^{36}$~erg~s$^{-1}$. However \cite{Giacconi01}
found that the average AGN spectrum of sources in the Chandra Deep
Field South (CDFS) was softer than the sources considered here.  Even
the faintest (hardest) group of sources in the CDFS are found to have
$\Gamma = 1.35 (\pm 0.20)$, which is softer than the interval $0.8 \le
\Gamma \le 1.2$ we focus on. We conclude that only few, if any, of the
HMXB candidates are AGN. All 18 of these sources should be followed up
with further optical observations to investigate the nature of the
donor star and confirm their status as HMXBs in M31.

\subsection{Luminosity Functions}
\label{subsec:CLF}

\begin{figure*}
\begin{center}
\includegraphics[height=12cm, angle=-90]{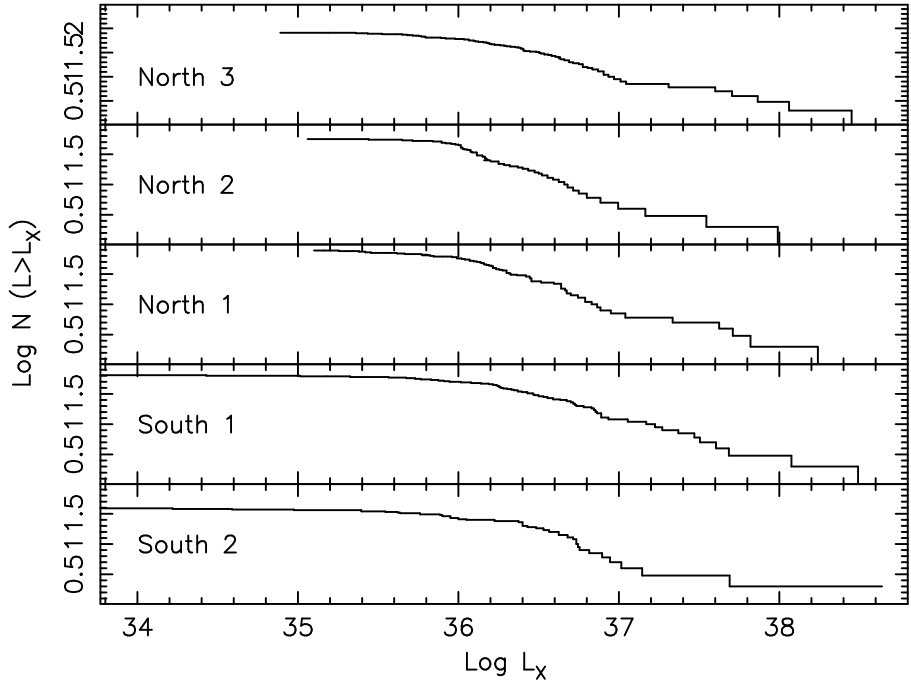}
\end{center}
\caption {Cumulative luminosity functions of each of the disc
  fields. The y-axes range is 0 to 2, except for North 3 where it is
  0-2.5.}
\label{fig:CLF}
\end{figure*}

Figure~\ref{fig:CLF} shows the cumulative luminosity functions (CLFs)
of each of the disc fields with the 0.3-10 keV luminosity, L$_{\mathrm
  X}$, plotted on the x-axis and the number of sources with a
luminosity higher than L$_{\mathrm X}$ on the y-axis. The confirmed
foreground star (source 182), AGN (source 27) and background galaxy
(source 112) have been removed from the relevant fields, however we
have not removed any of the 59 sources which are only classified as
foreground or background candidates by \cite{PFH05}.  The change
between freely fit and faint sources in each field occurs between
$5\times10^{35}$ erg s$^{-1}$ and $1.7\times10^{36}$ erg s$^{-1}$.
The source luminosities below this limit have been calculated using
the conversion factor from the summed faint source spectrum (see Table
\ref{tab:pimms}).

South 2 and North 2 have the highest luminosity cutoffs of freely fit
sources. This is due to these observations having the shortest good
time due to background flaring.  These regions also have the smallest
number of point source detections. South 1 and South 2 have two and
three sources respectively with luminosities below $1\times10^{35}$
erg s$^{-1}$, while the other fields only have sources above this
limit. South 1, North 1 and North 3 have similar numbers of point
source detections and the change between freely modelled sources and
faint sources occurs at a similar luminosity.

\begin{figure}
\begin{center}
\includegraphics[scale=0.3, angle=-90]{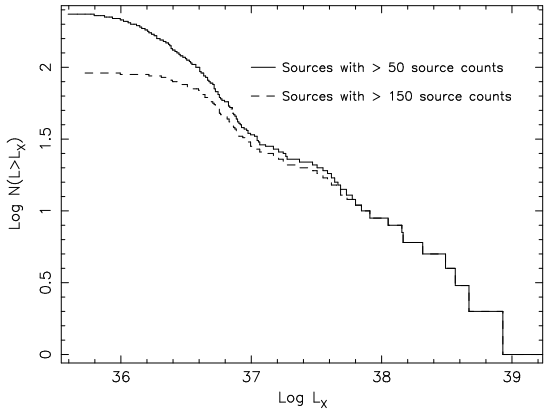}
\end{center}
\caption{Cumulative luminosity functions of sources with more than 50
  source counts (solid) and of sources with more than 150 source
  counts (dashed).}
\label{fig:50vs150}
\end{figure}

In Fig.~\ref{fig:50vs150} we present the CLF of all the disc sources
combined, for sources with more than 50 source counts, and compare it
to the CLF of sources with more than 150 source counts.  For both
these CLFs we have removed the three sources known not to belong to
the disc of M31. This comparison is in order to check the validity of
results derived from the spectral fitting of sources with only 50
source counts. Above $\sim10^{37}$ erg s$^{-1}$ the two functions are
almost identical. Using
Sherpa\footnote{http://cxc.harvard.edu/sherpa/index.html}, a straight
line fit of the CLFs above 10$^{36}$ erg s$^{-1}$ in
Fig.~\ref{fig:50vs150} gives a slope $\alpha = 0.7$ for the 50 counts
case and $\alpha = 0.6$ for the 150 counts case.  The flattening of
the CLF at $\sim10^{37}$ erg s$^{-1}$ is clearly apparent in the CLF
of sources with more than 150 source counts, hence demonstrating that
it is not an artifact of low count rate source fitting.

Previously \citet{Williams04}, using \emph{Chandra}, found that the
northern disc had fewer sources above $10^{37}$ erg s$^{-1}$ than the
southern disc. They found 10 sources with luminosities this value in
the southern disc and only 5 in the northern disc, while sources with
luminosities below this value are more evenly distributed, with 12 in
the southern disc and 11 in the northern \citep[numbers from
Fig.~11,][]{Williams04}. We find that South 1 may be over abundant in
bright sources with 11 non globular cluster sources above $10^{37}$
erg s$^{-1}$, while the other fields have somewhat smaller
numbers. There are 5 bright non globular cluster sources in North 1, 3
in North 2, 7 in North 3 and 4 in South 2.  In total we find 15 non
globular cluster sources brighter than $10^{37}$ erg s$^{-1}$ in the
southern disc and 15 in the northern disc. As there are three northern
disc fields and two southern the average number of bright sources per
field in the southern disc is slightly larger than in the northern
disc; however the difference in the number of sources is not as
pronounced as that seen by \cite{Williams04}.  This difference is
consistent with the discussion below relating to the comparison
between using individual spectral fitting and using a single simple
spectral model as assumed by \citet{Williams04}

\begin{figure}
\begin{center}
\includegraphics[scale=0.3, angle=-90]{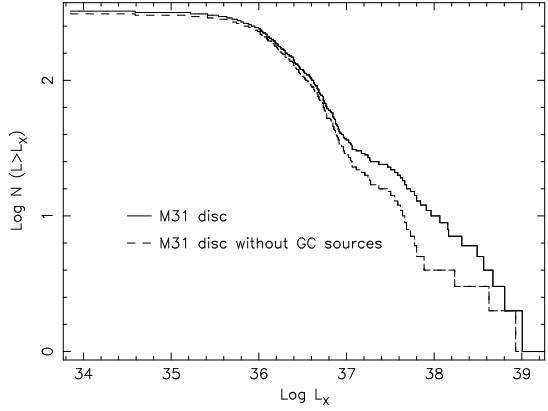}
\end{center}
\caption{Cumulative luminosity function of all source in the disc of
  M31 (solid), compared to the luminosity function of the disc with
  known globular cluster sources removed (dashed).}
\label{fig:whole_CLF}
\end{figure}

We identify the luminosity below which these observations are
incomplete as the luminosity at which we see a break in the CLF of the
whole disc, see Fig.~\ref{fig:whole_CLF}. This limit is $\sim
1\times10^{36}$ erg s$^{-1}$ which is in line with limits quoted in
\citet{Trud02} (detection limit ~$5\times10^{35}$ erg s$^{-1}$,
completeness limit $\sim10^{36}$ erg s$^{-1}$ for the central region,
North 1 and North 2).  \emph{Chandra} surveys (\citet{Kong03}
(detection limit ~$10^{35}$ erg s$^{-1}$, completeness limit
~$10^{36}$ erg s$^{-1}$) and \citet{Williams04} (completeness limit
~$4\times10^{36}$ erg s$^{-1}$ in the disc)) also have similar limits.

The CLF of each field was fit individually with a power law above and
below the completeness break L$_\mathrm{b} = 1\times10^{36}$ erg
s$^{-1}$. The results are given in Table \ref{tab:CLF} (columns 2 \& 3
entitled ``Freely Fit'').  The slopes of the disc CLFs above
L$_\mathrm{b}$ are between the values expected for starburst galaxies
and for spiral galaxies from \citet{Kilgard02}.  Given that we are
examining the disc of a spiral galaxy, this is to be expected.
Fitting the CLF of the bulge of M31, \citet{Shirey01} find a slope of
$1.77\pm0.35$ for $37.4\leq$~log~L$_{\mathrm X}$~erg~s$^{-1}~<38.1$,
flattening to $\alpha=0.43$ for log L$_{\mathrm X}$ erg s$^{-1}
<37.4$. According to the surveys of \citet{Colbert04} and
\citet{Kilgard02} galaxies with ongoing or recent star formation show
flatter CLFs than elliptical galaxies consisting of old populations.
Comparing the slope of the bulge CLF from \citet{Shirey01} with those
of the disc from this work (see Table \ref{tab:CLF}) show that the CLF
of the disc is flatter than that of the core.  This result is
consistent with the fact that there is more on-going star formation in
the disc of M31 than in the core.

\begin{table*}
  \begin{center}
    \caption{Field by field comparison of the slopes above and below
      the break luminsoty (L$_\mathrm{b}$) for the CLFs created from
      freely fit spectra and the CLFs of luminosities derived from
      fixed models. L$_\mathrm{b}$ is set to 10$^{36}$ erg s$^{-1}$,
      the slopes are for the CLF above and below L$_\mathrm{b}$. Data
      for the bulge of M31 comes from \cite{Shirey01} where the break
      is $2.5 \times 10^{36}$ erg~s$^{-1}$.}
    \label{tab:CLF}
    \begin{tabular}{c c c c c}
      \noalign{\smallskip}
      \hline
      \hline
      \noalign{\smallskip}
      Data & \multicolumn{2}{c}{Freely Fit} & \multicolumn{2}{c}{Fixed Model} \\
        & CLF slope above L$_\mathrm{b}$ & CLF slope below L$_\mathrm{b}$ & CLF slope above L$_\mathrm{b}$ & CLF slope below L$_\mathrm{b}$ \\
        \noalign{\smallskip}
        \hline
        \noalign{\smallskip}
        North 3 & 0.7 & 0.15 & 0.8 & 0.4 \\
        North 2 & 0.8 & 0.12 & 0.8 & 0.5 \\
        North 1 & 0.7 & 0.13 & 0.8 & 0.8 \\
        South 1 & 0.7 & 0.05 & 1.0 & 0.2 \\
        South 2 & 0.5 & 0.06 & 0.6 & 0.4 \\
        Bulge & 1.8(4) & - \\ 
      \noalign{\smallskip}
      \hline
      \noalign{\smallskip}
    \end{tabular}
  \end{center}
\end{table*}

\begin{figure}
\begin{center}
\includegraphics[scale=0.3, angle=-90]{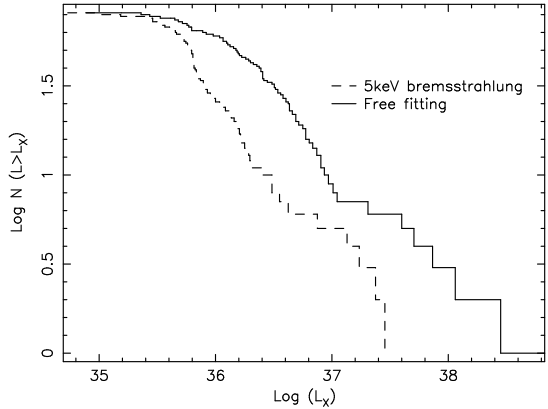}
\end{center}
\caption{Cumulative luminosity functions of North 3. The solid line
  consists of both the freely fit and faint sources from this work,
  while the dashed line is the luminosity of sources derived from
  assuming a bremsstrahlung emission model with k$T$~=~5~keV
  \citep{Kilgard02}.}
\label{fig:kil_CLF}
\end{figure}

\citet{Kilgard02} analysed seven spiral and starburst galaxies, not
including M31, in a \emph{Chandra} mini-survey.  They used an absorbed
5~keV bremsstrahlung emission model to convert from count rate to flux
for all detected sources. In order to make a direct comparison between
our work and the results from \citet{Kilgard02}, we used a 5~keV
thermal bremsstrahlung model with fixed photoelectric absorption
(n$_\mathrm{H} = 0.1 \times 10^{22}$~H~atom~cm$^{-2}$) to calculate
the luminosity of all the M31 disc sources.  Figure \ref{fig:kil_CLF}
shows the North 3 CLF obtained in this way, as well as the freely fit
CLF.  There are two main differences: firstly, fixing the model gave
sources which were fainter on the whole than the freely fit sources,
and the total luminosity of each field was reduced to only
$1-3\times10^{38}$~erg~s$^{-1}$ rather than
$4-20\times10^{38}$~erg~s$^{-1}$. The second effect was the change in
average slope of the CLF of each of the fields from $\alpha \simeq
0.7$ when the source spectra were freely fit, to $\alpha \simeq 0.8$
for the fixed model sources.  \citet{Kilgard02} conclude that steeper
slopes of CLFs imply less star formation. We have found that the
freely fit CLF was not as steep as the CLF of the fixed model sources
and thus the CLF slope-SFR calibration could be systematically offset.
A similar finding for NGC 253 is discussed in detail by
\citet{Barnard07conf} and \citet{Barnard07253}.

Figure ~\ref{fig:whole_CLF} shows the CLF of all the disc sources
(excluding the 3 identified foreground and background objects), with
and without globular cluster (GC) sources. It can be seen that there
is a distinct flattening of the CLF in the range $37.0 \le \log
\mathrm{L_X /erg s^{-1}} \le 37.5$ present in both samples. A KS test
shows that the sample with GC sources has a 2.5\% chance being drawn
from the same population that is represented by the best fit power law
of the data between $36 \le \log \mathrm{L_X /erg s^{-1}} \le 38$,
while the corresponding probability for the sample without GC sources
is 2.1\%. The K-S probability becomes larger if a power law fit for
the full range of luminosities, including the 8 and 3 sources,
respectively, that are brighter than $10^{38}$~erg~s$^{-1}$, is
considered.  Although the KS test remains inconclusive we point out
that the reality of the flattening in the CLF is supported by the fact
that it can also be seen in the CLFs of the M31 disc sources derived
from prescribed models (see Fig.~\ref{fig:kil_CLF}), in the higher
count rate sources only (see Fig.~\ref{fig:50vs150}) and in some of
the individual fields of M31 (especially North 3, North 1 and South 2,
see Fig.~\ref{fig:CLF}).  As well as in the CLF of M31 sources, it is
possible that the same feature is also present in the CLFs of the SMC
\citep{SMC05} and M33 \citep{Grimm05} which are both predominantly
young populations but have some evidence for a LMXB contribution.

This feature could be due to the emission from a mixture of HMXB and
LMXB populations \citep[see][for work on the sub-populations of the
Milky Way]{Grimm02} or possibly due to a change in the nature of the
compact object. Kalogera (2007) have reported that they have observed
such a dip in theoretical population models, associated with the
transition from a binary population with main sequence donors (below
$\sim10^{37}$ erg s$^{-1}$) to a population with red giant donors
(above the dip). This is because, for most magnetic braking laws, the
mass transfer rate driven by nuclear expansion of donors (as in red
giant donors) is higher than that for mass transfer driven by orbital
angular momentum losses (short period systems with main sequence
donors). We show in Sec~\ref{subsec:contamination} below that the
flattening is not due to a change in the background AGN CLF.

\begin{figure}
\begin{center}
\includegraphics[scale=0.3, angle=-90]{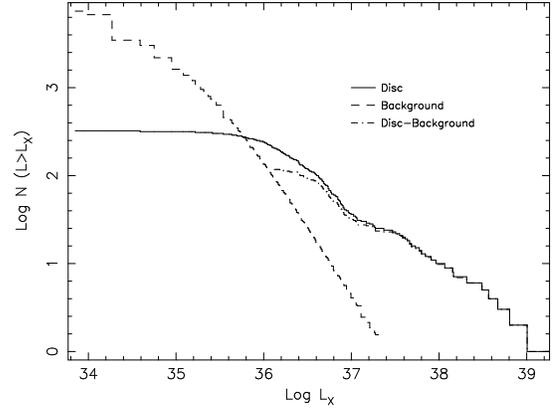}
\end{center}
\caption{Cumulative luminosity functions of the population of sources
  in the disc of M31 from this work (solid) and the derived AGN
  background (dashed).  Also shown for luminosities above
  1.4$\times10^{36}$ erg s$^{-1}$ is the CLF of the disc with the
  derived background contribution subtracted (dot-dash).}
\label{fig:Background_CLF}
\end{figure}

\subsection{AGN Contamination}
\label{subsec:contamination}
The CLF for the combined disc flattens below $1\times10^{36}$ erg
s$^{-1}$, while the very faintest sources are around $10^{34}$ erg
s$^{-1}$.  We expect that there is a significant contribution from
background AGN at these faint luminosities.

\citet{Moretti03} consider the AGN contribution in a hard (2-10~keV)
and soft (1-2~keV) energy band.  We assume a standard AGN spectrum, a
power law model with $\Gamma = 1.44$ \citep[following][]{Rosati02}, to
calculate the AGN flux in our 0.3-10~keV band. We calculated the
number of background sources in each field above the 0.3-10~keV
incompleteness limit of L$_\mathrm{X} = 10^{36}$ erg s$^{-1}$
(corresponding to a 1-2~keV flux limit of
$8.5\times10^{-14}$~erg~s$^{-1}$ and a 2-10~keV limit of
$1.9\times10^{-14}$~erg~s$^{-1}$). Given that the area of each field
is 0.20 deg$^2$ and using \citet[Eq.~2]{Moretti03} we find that there
are 26-30 background sources above $10^{36}$ erg s$^{-1}$ in each disc
field of M31.  The lower limit is calculated assuming that all the
sources visible in the soft band are also seen in the hard band, the
upper limit assumes that none of the sources seen in the soft band are
seen in the hard band.

We can then estimate the shape of the CLF intrinsic to the M31 disc by
removing the background AGN contribution according to
\citet{Moretti03}. In Fig.~\ref{fig:Background_CLF} we show the
observed complete disc CLF (Disc), with a total area of 0.98 deg$^2$,
and the calculated sum of the hard and soft background contributions
per 0.98 deg$^2$ (Background). According to \citet{Moretti03} the soft
sources do not contribute above $\sim2.8\times10^{36}$ erg s$^{-1}$
and so the total background contribution is very close to the number
of sources seen in the hard (2-10 keV) band. The upper limit total is
shown in Fig.~\ref{fig:Background_CLF} as a dashed line. The model
shows that the observations are incomplete below a
few$\times10^{36}$. We also show our estimation of the CLF of sources
intrinsic to M31 (Disc CLF with the Background CLF subtracted) only
above the luminosity at which there are more sources in the disc than
in the background ($1.4\times10^{36}$ erg s$^{-1}$). Below this
luminosity the complete M31 CLF is dominated by the backgroud
contribution, and the incompleteness of the survey is obvious. We note
that the flattening of the CLF near 10$^{37}$ erg s$^{-1}$ is still
very prominent in the background corrected CLF.

\section{Conclusions}
\label{sec:conclusion}
We have revisited five archival XMM-Newton observations of the disc of
M31.  These data revealed 335 point detections across the 5 fields
constituting 329 discrete point sources. All the sources were fit with
three spectral models: blackbody, bremsstrahlung and power law and the
results of these fits were examined.

Using only the best fit power law model to each source, we
investigated the distribution of photon indices of these fits. The
broad range of photon index seen in Fig.~\ref{fig:histo} and the
difference in the CLFs seen in Fig.~\ref{fig:kil_CLF} cast doubt on
the validity of assuming the same spectral model for all sources when
analysing more distant X-ray point source populations.  Individual
spectral fitting has identified the first 18 HMXB candidates in the
disc.  The HMXB candidates are all best fit by a power law with a
photon index of 0.8-1.2 indicating magnetically acreting neutron
stars. 

For the first time X-ray point sources in M31 with as few as 50 source
counts have been individually spectrally fit, in contrast to previous
surveys which have used the same assumed model for all sources except
the very brightest. This has led us to create the first CLFs of the
M31 disc region created from 240 individually spectrally fit sources
shown in Fig.~\ref{fig:CLF}. It can be seen that the CLFs of the
fields are quite similar across the disc of M31 and that there are no
obvious changes in the CLF slopes with increasing apparent distance
from the core.  The CLFs of both individual fields
(Fig.~\ref{fig:CLF}) and the entire disc both with and without GC
sources (Fig.~\ref{fig:whole_CLF}) show a distinct flattening between
37.0 $\lesssim \log \mathrm{L_X}$ erg s$^{-1} \lesssim 37.5$.  This
flattening could be due to the emission from a mixture of HMXB and
LMXB populations or due to a change in the nature of the compact
object or the donor star. This prominent flattening in the CLF of M31
sources may also appear at a lower statistical significance in the
CLFs of several other galaxies.

Each observation contains not only the point sources in M31 but also
some contamination from foreground and background sources. We have
estimated that there are around 20-31 background AGN above $10^{36}$
erg s$^{-1}$, in each field observed, and find that above
$1.4\times10^{36}$ erg s$^{-1}$ there are few background AGN according
to \citet{Moretti03}. The CLF here is dominated by the sources
intrinsic to the disc of M31 and any foreground interlopers. Following
the correlations of the PSPC ROSAT survey with optical catalogues
\citep{Supper01}, we expect 5-10 forground objects above $10^{36}$ erg
s$^{-1}$ for each XMM-Newton field \citep{Shirey01}.

M31 is a prime target for population surveys because of its proximity
and similarity to our own Galaxy. We have investigated the X-ray point
sources in the disc of M31 in detail and this population challenges
theoretical models to explain the features seen in the distribution of
sources and in the CLF.

\begin{acknowledgements} 
  We would like to thank the referee, Sergei Trudolyubov for very
  helpful comments. We also thank Wolfgang Pietsch and Simon J. Clark
  for useful conversations during this work. LSG acknowledges support
  from the Open University. Astronomy research at the Open University
  is supported by a STFC rolling grant. JPO acknowledges support from
  STFC.

\end{acknowledgements}
\bibliographystyle{aa}
\bibliography{m31_13Aug_astroph}

\Online
\begin{appendix}
\section{}
In this section we give positional and spectral information on all 335
point source detections in five archival XMM-Newton observations of
the disc of M31. Table \ref{tab:sources} gives the source numbers in
order of RA and the positions as returned by the source detection
routines (i.e. no astrometric corrections were applied).  We also
provide details of which camera the source spectra or counts are taken
from and details of the best fit spectral model for each source,
including the absorption (n$_\mathrm{H}$), best fit photon index
($\Gamma$) or temperature (k$T$) depending on which model, and the
luminosity derived from that fit.  The symbol f signifies that the
absorption was fixed to $0.1 \times 10^{22}$~H~atom~cm$^{-2}$ for that
source. Note that several of the faint sources display a very large
formal uncertainty in n$_\mathrm{H}$.  The actualy best fit values
were accepted only if they exceeded the Galactic foreground value
(n$_\mathrm{H} = 0.1 \times 10^{22}$~H~atom~cm$^{-2}$). Errors on the
absorption and $\Gamma$/k$T$ are the two sided, non-symmetric errors
derived by xspec. The luminosity given is the mean value of the 90\%
confidence interval and hence has symetric errors quoted in
brackets. Best fit values are given for all sources with more than 50
source counts; even for cases where the fit implies significant or
very large error bars. Sources with less than 50 source counts are
denoted with ``faint'' as their best fit model, these sources are then
summed by field and the parameters of the best fit power law to the
summed spectrum applied to each source. Hence for the faint sources no
absorption or $\Gamma$/k$T$ are given.  Finally we give a classification
for each source.  These are the classifications and source number from
\citet{PFH05} where the sources appear in that work.  For the 34
sources not in \citet{PFH05} as well as source 238 we give our own
classifications and each of these have a $^1$ next to it.

\begin{center}
\renewcommand{\arraystretch}{1.5}
\onecolumn
\begin{landscape}

  \begin{longtable}{ccccccccc}
    \caption{Position and spectral properties of each source detected
      in five archival XMM-Newton observations of the disc of M31. We
      give the detected source position and detection camera(s) for
      each source.  The best fit model to a source can be faint (less
      than 50 source counts), pl (power law), bb (blackbody), br
      (bremsstrahlung), nsa (neutron star atmosphere) or diskbb (disk
      blackbody). Unless a source is faint we then give the absorption
      (n$_\mathrm{H}$) and appropriate parameter (photon index,
      $\Gamma$, or temperature, k$T$) of the best fit model. The symbol
      f signifies that the absorption was fixed to $0.1 \times
      10^{22}$~H~atom~cm$^{-2}$ for that source. For all sources we
      then quote the luminosity for that source derived from the model
      parameters with 90\% confidence errors in brackets, and finally
      a classification, either from \citet{PFH05} with the source
      number, or this work (denoted with a
      $^1$). \label{tab:sources}}\\ \hline\hline \noalign{\smallskip}

    Source & RA & Dec & Camera & Best Fit & n$_\mathrm{H}$ & $\Gamma$/k$T$ &
Luminosity & Classification \\ & (J2000) & (J2000) & & Model &
/$10^{22}$ cm$^{-2}$ & keV & /$10^{36}$ erg s$^{-1}$ &
\\ 

\noalign{\smallskip} \hline \noalign{\smallskip} \endfirsthead
    \caption{continued.}\\
    \hline\hline
    \noalign{\smallskip}
    Source & RA & Dec & Camera & Best Fit & n$_\mathrm{H}$ & $\Gamma$/k$T$ & Luminosity & Classification \\
    & (J2000) & (J2000) & & Model & /$10^{22}$ cm$^{-2}$ & keV & /$10^{36}$ erg s$^{-1}$ & \\
    \noalign{\smallskip}
    \hline
    \noalign{\smallskip}
    \endhead
    \hline
    \endfoot
1	&	00	:	38	:	56.5	&	+	40	:	34	:	51	&		&	faint	&							&							&	-	&	$\mathrm{	<hard>	1		}$	\\
2	&	00	:	38	:	59.9	&	+	40	:	39	:	11	&		&	faint	&							&							&	0.044	&	$\mathrm{	<fg star>	2		}$	\\
3	&	00	:	39	:	23.8	&	+	40	:	29	:	56	&	mos	&	pl	&	0.3	$_{-	0.3	}^{+	3.0	}$	&	1.1	$_{-	1.4	}^{+	3.8	}$	&	2.85 (2.01)	&	$\mathrm{	<hard>	11		}$	\\
4	&	00	:	39	:	25.1	&	+	40	:	37	:	20	&		&	faint	&							&							&	0.300	&	$\mathrm{			<hard>^1	}$	\\
5	&	00	:	39	:	27.3	&	+	40	:	46	:	47	&		&	faint	&							&							&	0.008	&	$\mathrm{	<hard>	15		}$	\\
6	&	00	:	39	:	29.0	&	+	40	:	35	:	42	&	pn \& mos	&	pl	&	0.16	$_{-	0.16	}^{+	1.13	}$	&	1.6	$_{-	0.8	}^{+	2.3	}$	&	1.23 (0.63) / 1.27 (0.68)	&	$\mathrm{	<hard>	18		}$	\\
7	&	00	:	39	:	31.6	&	+	40	:	36	:	16	&		&	faint	&							&							&	0.821	&	$\mathrm{	<hard>	19		}$	\\
8	&	00	:	39	:	36.6	&	+	40	:	35	:	29	&		&	faint	&							&							&	0.767	&	$\mathrm{	<hard>	22		}$	\\
9	&	00	:	39	:	38.7	&	+	40	:	47	:	57	&	pn	&	pl	&	f						&	0.8	$_{-	0.7	}^{+	0.9	}$	&	5.40 (3.83)	&	$\mathrm{	<hard>	23		}$	\\
10	&	00	:	39	:	40.3	&	+	40	:	35	:	31	&	pn \& mos	&	pl	&	0.24	$_{-	0.12	}^{+	0.18	}$	&	0.6	$_{-	0.3	}^{+	0.2	}$	&	3.87 (0.95) / 3.85 (0.78)	&	$\mathrm{	<hard>	24		}$	\\
11	&	00	:	39	:	43.5	&	+	40	:	39	:	44	&	pn	&	bb	&	f						&	0.21	$_{-	0.04	}^{+	0.05	}$	&	1.02 (0.5)	&	$\mathrm{	<fg star>	26		}$	\\
12	&	00	:	39	:	45.7	&	+	40	:	44	:	54	&	pn	&	pl	&	1.9	$_{-	1.9	}^{+	8.5	}$	&	1.6	$_{-	1.6	}^{+	3.8	}$	&	3.47 (3.19)	&	$\mathrm{	<hard>	28		}$	\\
13	&	00	:	39	:	47.9	&	+	40	:	34	:	35	&		&	faint	&							&							&	0.816	&	$\mathrm{	<AGN>	29		}$	\\
14	&	00	:	39	:	48.9	&	+	40	:	35	:	14	&	pn \& mos	&	pl	&	0.7	$_{-	0.2	}^{+	0.4	}$	&	2.3	$_{-	0.5	}^{+	0.7	}$	&	3.32 (1.21) / 3.19 (1.05)	&	$\mathrm{	<hard>	30		}$	\\
15	&	00	:	39	:	56.3	&	+	40	:	41	:	00	&	pn \& mos	&	nsa	&	1.07	$_{-	0.17	}^{+	0.2	}$	&	5.36	$_{-	0.03	}^{+	0.02	}$	&	1350 (1020) / 1250 (910)	&	$\mathrm{	<fg star>	31		}$	\\
16	&	00	:	39	:	57.9	&	+	40	:	27	:	26	&	pn	&	bb	&	0.26	$_{-	0.22	}^{+	0.16	}$	&	0.13	$_{-	0.02	}^{+	0.06	}$	&	5.57 (3.23)	&	$\mathrm{	<SNR>	32		}$	\\
17	&	00	:	39	:	59.7	&	+	40	:	31	:	59	&	pn \& mos	&	pl	&	0.28	$_{-	0.04	}^{+	0.08	}$	&	2.3	$_{-	0.2	}^{+	0.2	}$	&	17.0 (3.1) / 19.4 (2.2)	&	$\mathrm{	<hard>	33		}$	\\
18	&	00	:	40	:	00.5	&	+	40	:	26	:	41	&		&	faint	&							&							&	0.202	&	$\mathrm{			<hard>^1	}$	\\
19	&	00	:	40	:	01.1	&	+	40	:	25	:	24	&	pn	&	pl	&	f						&	1.2	$_{-	0.5	}^{+	0.6	}$	&	2.59 (1.56)	&	$\mathrm{	<hard>	34		}$	\\
20	&	00	:	40	:	01.4	&	+	40	:	33	:	23	&	pn	&	pl	&	0.9	$_{-	0.4	}^{+	0.9	}$	&	3.2	$_{-	1.0	}^{+	2.3	}$	&	4.60 (3.29)	&	$\mathrm{	<hard>	36		}$	\\
21	&	00	:	40	:	01.6	&	+	40	:	32	:	43	&	pn \& mos	&	pl	&	0.4	$_{-	0.3	}^{+	0.6	}$	&	0.9	$_{-	0.6	}^{+	0.4	}$	&	2.27 (0.97) / 2.66 (1.12)	&	$\mathrm{	<hard>	35		}$	\\
22	&	00	:	40	:	06.1	&	+	40	:	24	:	09	&		&	faint	&							&							&	-	&	$\mathrm{	<hard>	37		}$	\\
23	&	00	:	40	:	06.6	&	+	40	:	21	:	48	&		&	faint	&							&							&	-	&	$\mathrm{			SNR^1	}$	\\
24	&	00	:	40	:	07.1	&	+	40	:	41	:	42	&		&	faint	&							&							&	0.783	&	$\mathrm{	<hard>	40		}$	\\
25	&	00	:	40	:	07.5	&	+	40	:	31	:	14	&	pn \& mos	&	pl	&	0.28	$_{-	0.12	}^{+	0.52	}$	&	3.2	$_{-	1.3	}^{+	3.3	}$	&	2.43 (1.63) / 1.68 (1.13)	&	$\mathrm{	<fg star>	42		}$	\\
26	&	00	:	40	:	08.6	&	+	40	:	47	:	12	&		&	faint	&							&							&	-	&	$\mathrm{	<hard>	43		}$	\\
27	&	00	:	40	:	13.8	&	+	40	:	50	:	06	&	pn	&	pl	&	0.41	$_{-	0.02	}^{+	0.02	}$	&	1.88	$_{-	0.05	}^{+	0.05	}$	&	753 (23)	&	$\mathrm{	AGN	50		}$	\\
28	&	00	:	40	:	13.9	&	+	40	:	35	:	33	&	pn \& mos	&	bb	&	0.7	$_{-	0.6	}^{+	0.4	}$	&	0.1	$_{-	0.04	}^{+	0.1	}$	&	7.36 (7.35) / 7.78 (7.75)	&	$\mathrm{	<fg star>	49		}$	\\
29	&	00	:	40	:	14.3	&	+	40	:	51	:	35	&		&	faint	&							&							&	-	&	$\mathrm{			<hard>^1	}$	\\
30	&	00	:	40	:	14.3	&	+	40	:	33	:	41	&	pn \& mos	&	pl	&	0.16	$_{-	0.05	}^{+	0.09	}$	&	1.38	$_{-	0.14	}^{+	0.13	}$	&	9.25 (1.26) / 9.47 (1.17)	&	$\mathrm{	<hard>	51		}$	\\
31	&	00	:	40	:	16.7	&	+	40	:	53	:	07	&		&	faint	&							&							&	-	&	$\mathrm{	<hard>	53		}$	\\
32	&	00	:	40	:	16.8	&	+	40	:	50	:	37	&	pn	&	pl	&	1.0	$_{-	1.0	}^{+	2.3	}$	&	1.7	$_{-	1.5	}^{+	2.4	}$	&	5.38 (3.95)	&	$\mathrm{			<hard>^1	}$	\\
33	&	00	:	40	:	17.8	&	+	40	:	32	:	57	&	pn	&	pl	&	0.3	$_{-	0.2	}^{+	3.8	}$	&	2.0	$_{-	1.1	}^{+	8.0	}$	&	0.98 (0.56)	&	$\mathrm{	<hard>	54		}$	\\
34	&	00	:	40	:	18.2	&	+	40	:	48	:	41	&	pn \& mos	&	pl	&	0.20	$_{-	0.15	}^{+	0.38	}$	&	1.2	$_{-	0.5	}^{+	0.7	}$	&	8.39 (3.87) / 7.35 (2.66)	&	$\mathrm{			<hard>^1	}$	\\
35	&	00	:	40	:	20.2	&	+	40	:	43	:	59	&	pn \& mos	&	pl	&	0.16	$_{-	0.02	}^{+	0.02	}$	&	1.58	$_{-	0.06	}^{+	0.06	}$	&	140 (15) / 225 (12)	&	$\mathrm{	GlC	55		}$	\\
36	&	00	:	40	:	20.9	&	+	40	:	39	:	18	&	pn \& mos	&	pl	&	0.8	$_{-	2.4	}^{+	1.3	}$	&	2	$_{-	1	}^{+	3	}$	&	2.44 (1.55) / 2.44 (1.66)	&	$\mathrm{	<hard>	56		}$	\\
37	&	00	:	40	:	22.6	&	+	40	:	36	:	09	&	pn \& mos	&	pl	&	0.25	$_{-	0.15	}^{+	0.12	}$	&	2.1	$_{-	0.4	}^{+	0.3	}$	&	5.05 (1.49) / 6.17 (1.14)	&	$\mathrm{	<hard>	58		}$	\\
38	&	00	:	40	:	23.6	&	+	40	:	53	:	05	&	mos	&	bb	&	f						&	0.20	$_{-	0.06	}^{+	0.07	}$	&	0.79 (0.56)	&	$\mathrm{	<fg star>	59		}$	\\
39	&	00	:	40	:	24.1	&	+	40	:	29	:	45	&	pn \& mos	&	pl	&	0.16	$_{-	0.07	}^{+	0.08	}$	&	1.7	$_{-	0.2	}^{+	0.2	}$	&	11.6 (1.6) / 12.6 (1.4)	&	$\mathrm{	<hard>	60		}$	\\
40	&	00	:	40	:	27.6	&	+	40	:	46	:	34	&		&	faint	&							&							&	0.407	&	$\mathrm{	<hard>	65		}$	\\
41	&	00	:	40	:	29.5	&	+	40	:	37	:	06	&	pn \& mos	&	pl	&	0.8	$_{-	0.7	}^{+	1.6	}$	&	3	$_{-	1	}^{+	2	}$	&	2.43 (1.61) / 3.23 (2.33)	&	$\mathrm{	<hard>	66		}$	\\
42	&	00	:	40	:	31.6	&	+	40	:	58	:	34	&		&	faint	&							&							&	0.153	&	$\mathrm{	<SNR>	70		}$	\\
43	&	00	:	40	:	32.6	&	+	41	:	00	:	45	&		&	faint	&							&							&	-	&	$\mathrm{	<hard>	71		}$	\\
44	&	00	:	40	:	33.2	&	+	40	:	49	:	39	&	pn	&	pl	&	0.3	$_{-	0.3	}^{+	1.7	}$	&	1.4	$_{-	1.2	}^{+	4.1	}$	&	2.08 (1.30)	&	$\mathrm{	<hard>	72		}$	\\
45	&	00	:	40	:	37.7	&	+	40	:	40	:	45	&		&	faint	&							&							&	-	&	$\mathrm{	<SSS>	75		}$	\\
46	&	00	:	40	:	39.6	&	+	41	:	06	:	10	&	mos	&	bb	&	0.6	$_{-	0.6	}^{+	2.7	}$	&	1.6	$_{-	0.3	}^{+	0.4	}$	&	1.58 (1.55)	&	$\mathrm{	<hard>	77		}$	\\
47	&	00	:	40	:	40.0	&	+	40	:	25	:	47	&		&	faint	&							&							&	0.431	&	$\mathrm{	<hard>	78		}$	\\
48	&	00	:	40	:	42.9	&	+	40	:	32	:	41	&	pn \& mos	&	pl	&	0.6	$_{-	0.3	}^{+	0.6	}$	&	2.7	$_{-	0.7	}^{+	1.5	}$	&	5.44 (2.69) / 5.83 (2.83)	&	$\mathrm{	<hard>	81		}$	\\
49	&	00	:	40	:	44.3	&	+	40	:	48	:	58	&	pn \& mos	&	pl	&	0.7	$_{-	0.2	}^{+	0.7	}$	&	1.5	$_{-	0.4	}^{+	0.7	}$	&	3.71 (1.06) / 4.14 (1.04)	&	$\mathrm{	<hard>	84		}$	\\
50	&	00	:	40	:	45.4	&	+	40	:	51	:	37	&	pn \& mos	&	bb	&	f						&	0.18	$_{-	0.02	}^{+	0.03	}$	&	0.75 (0.37) / 0.85 (0.43)	&	$\mathrm{	<hard>	88		}$	\\
51	&	00	:	40	:	46.8	&	+	40	:	29	:	12	&		&	faint	&							&							&	0.441	&	$\mathrm{	<fg star>	90		}$	\\
52	&	00	:	40	:	47.1	&	+	40	:	55	:	20	&	pn \& mos	&	pl	&	0.32	$_{-	0.18	}^{+	0.18	}$	&	2.2	$_{-	0.6	}^{+	0.8	}$	&	1.86 (0.67) / 1.45 (0.510)	&	$\mathrm{	SNR	91		}$	\\
53	&	00	:	40	:	48.1	&	+	40	:	51	:	11	&	pn	&	pl	&	f						&	2.7	$_{-	1.6	}^{+	1.8	}$	&	1.95 (1.43)	&	$\mathrm{	<hard>	92		}$	\\
54	&	00	:	40	:	48.8	&	+	40	:	49	:	25	&	mos	&	br	&	f						&	0.36	$_{-	0.11	}^{+	0.23	}$	&	1.27 (1.00)	&	$\mathrm{	<hard>	94		}$	\\
55	&	00	:	40	:	48.9	&	+	40	:	30	:	33	&	pn \& mos	&	pl	&	0.41	$_{-	0.18	}^{+	0.33	}$	&	2.1	$_{-	0.5	}^{+	0.8	}$	&	5.70 (1.87) / 3.09 (1.86)	&	$\mathrm{	<hard>	93		}$	\\
56	&	00	:	40	:	50.0	&	+	41	:	07	:	30	&		&	faint	&							&							&	0.495	&	$\mathrm{	<hard>	96		}$	\\
57	&	00	:	40	:	52.7	&	+	40	:	36	:	21	&		&	faint	&							&							&	0.035	&	$\mathrm{	<hard>	98		}$	\\
58	&	00	:	40	:	56.9	&	+	40	:	56	:	38	&	pn \& mos	&	pl	&	0.29	$_{-	0.12	}^{+	0.13	}$	&	2.2	$_{-	0.5	}^{+	0.7	}$	&	1.78 (0.56) / 2.02 (0.52)	&	$\mathrm{	<fg star>	101		}$	\\
59	&	00	:	41	:	00.3	&	+	41	:	00	:	26	&		&	faint	&							&							&	0.072	&	$\mathrm{	<hard>	105		}$	\\
60	&	00	:	41	:	06.4	&	+	40	:	27	:	08	&	mos	&	bb	&	f						&	0.49	$_{-	0.11	}^{+	0.15	}$	&	3.86 (2.09)	&	$\mathrm{	<hard>	108		}$	\\
61	&	00	:	41	:	07.6	&	+	40	:	50	:	47	&	pn	&	pl	&	0.3	$_{-	0.3	}^{+	2.5	}$	&	1.1	$_{-	0.7	}^{+	3.9	}$	&	1.83 (1.33)	&	$\mathrm{	<hard>	111		}$	\\
62	&	00	:	41	:	08.4	&	+	40	:	51	:	28	&	pn \& mos	&	pl	&	0.20	$_{-	0.09	}^{+	0.13	}$	&	1.6	$_{-	0.2	}^{+	0.3	}$	&	7.18 (1.63) / 7.53 (1.12)	&	$\mathrm{	<hard>	113		}$	\\
63	&	00	:	41	:	09.9	&	+	41	:	04	:	52	&	pn \& mos	&	pl	&	2.9	$_{-	1.3	}^{+	2.1	}$	&	2.6	$_{-	0.5	}^{+	1.3	}$	&	5.24 (3.9) / 4.85 (3.25)	&	$\mathrm{	<hard>	115		}$	\\
64	&	00	:	41	:	11.8	&	+	40	:	54	:	20	&	mos	&	bb	&	f						&	0.7	$_{-	0.2	}^{+	0.2	}$	&	0.74 (0.56)	&	$\mathrm{	<hard>	117		}$	\\
65	&	00	:	41	:	13.0	&	+	40	:	51	:	33	&		&	faint	&							&							&	-	&	$\mathrm{	<hard>	118		}$	\\
66	&	00	:	41	:	13.2	&	+	40	:	59	:	47	&	pn \& mos	&	pl	&	0.15	$_{-	0.05	}^{+	0.04	}$	&	2.18	$_{-	0.11	}^{+	0.16	}$	&	7.12 (1.07) / 7.73 (0.82)	&	$\mathrm{	<hard>	119		}$	\\
67	&	00	:	41	:	14.3	&	+	41	:	09	:	05	&		&	faint	&							&							&	0.549	&	$\mathrm{			<hard>^1	}$	\\
68	&	00	:	41	:	15.1	&	+	41	:	01	:	00	&	pn \& mos	&	br	&	0.39	$_{-	0.07	}^{+	0.03	}$	&	0.27	$_{-	0.04	}^{+	0.04	}$	&	47.5 (13.7) / 43.9 (13.8)	&	$\mathrm{	<hard>	122		}$	\\
69	&	00	:	41	:	18.1	&	+	41	:	06	:	43	&	pn \& mos	&	pl	&	0.21	$_{-	0.14	}^{+	0.10	}$	&	2.2	$_{-	0.4	}^{+	0.5	}$	&	1.74 (0.52) / 1.71 (0.44)	&	$\mathrm{	<hard>	125		}$	\\
70	&	00	:	41	:	18.5	&	+	40	:	51	:	58	&	pn \& mos	&	pl	&	0.13	$_{-	0.13	}^{+	0.23	}$	&	1.2	$_{-	0.5	}^{+	0.3	}$	&	2.89 (2.02) / 3.91 (2.10)	&	$\mathrm{	<fg star>	128		}$	\\
71	&	00	:	41	:	19.6	&	+	41	:	00	:	09	&		&	faint	&							&							&	0.147	&	$\mathrm{	<hard>	130		}$	\\
72	&	00	:	41	:	21.0	&	+	41	:	03	:	39	&	pn	&	pl	&	0.7	$_{-	0.7	}^{+	3.5	}$	&	1.6	$_{-	2.1	}^{+	8.4	}$	&	0.93 (0.75)	&	$\mathrm{	<hard>	132		}$	\\
73	&	00	:	41	:	21.5	&	+	41	:	07	:	54	&	pn \& mos	&	pl	&	0.17	$_{-	0.02	}^{+	0.02	}$	&	1.65	$_{-	0.07	}^{+	0.07	}$	&	30.1 (1.98) / 31.0 (1.62)	&	$\mathrm{	<hard>	134		}$	\\
74	&	00	:	41	:	25.2	&	+	40	:	51	:	11	&	pn \& mos	&	pl	&	f						&	2.07	$_{-	0.09	}^{+	0.09	}$	&	7.36 (0.88) / 7.93 (0.70)	&	$\mathrm{	<hard>	140		}$	\\
75	&	00	:	41	:	25.6	&	+	40	:	58	:	44	&	pn \& mos	&	pl	&	0.5	$_{-	0.3	}^{+	0.5	}$	&	1.5	$_{-	0.4	}^{+	0.6	}$	&	2.51 (0.79) / 2.48 (0.73)	&	$\mathrm{	<hard>	141		}$	\\
76	&	00	:	41	:	26.1	&	+	40	:	53	:	25	&	pn \& mos	&	pl	&	0.12	$_{-	0.06	}^{+	0.06	}$	&	1.7	$_{-	0.2	}^{+	0.2	}$	&	8.12 (1.44) / 7.97 (1.08)	&	$\mathrm{	<hard>	142		}$	\\
77	&	00	:	41	:	28.5	&	+	40	:	54	:	51	&		&	faint	&							&							&	0.615	&	$\mathrm{	<hard>	144		}$	\\
78	&	00	:	41	:	28.8	&	+	41	:	02	:	08	&	pn	&	nsa	&	0.28						&	5.96						&	0.84 (0.84)	&	$\mathrm{	<hard>	145		}$	\\
79	&	00	:	41	:	31.2	&	+	40	:	59	:	57	&	pn \& mos	&	pl	&	f						&	2.32	$_{-	0.16	}^{+	0.17	}$	&	3.33 (0.74) / 3.74 (0.61)	&	$\mathrm{	<hard>	147		}$	\\
80	&	00	:	41	:	35.5	&	+	41	:	06	:	53	&	pn \& mos	&	pl	&	0.14	$_{-	0.06	}^{+	0.05	}$	&	1.92	$_{-	0.19	}^{+	0.16	}$	&	6.63 (1.17) / 6.97 (1.00)	&	$\mathrm{	SNR	154		}$	\\
81	&	00	:	41	:	36.4	&	+	41	:	00	:	18	&	pn \& mos	&	pl	&	0.17	$_{-	0.13	}^{+	0.17	}$	&	1.6	$_{-	0.4	}^{+	0.3	}$	&	3.00 (0.97) / 2.50 (0.69)	&	$\mathrm{	<fg star>	157		}$	\\
82	&	00	:	41	:	37.7	&	+	41	:	01	:	07	&	pn \& mos	&	bb	&	1.00	$_{-	0.30	}^{+	0.07	}$	&	0.08	$_{-	0.01	}^{+	0.02	}$	&	328 (305) / 515 (474)	&	$\mathrm{	<hard>	159		}$	\\
83	&	00	:	41	:	41.0	&	+	41	:	03	:	33	&		&	faint	&							&							&	-	&	$\mathrm{	<AGN>	164		}$	\\
84	&	00	:	41	:	41.8	&	+	41	:	00	:	15	&		&	faint	&							&							&	-	&	$\mathrm{	<hard>	167		}$	\\
85	&	00	:	41	:	43.2	&	+	41	:	05	:	05	&	mos	&	pl	&	0.17	$_{-	0.17	}^{+	0.22	}$	&	1.9	$_{-	0.6	}^{+	0.8	}$	&	4.11 (0.14)	&	$\mathrm{	<fg star>	168		}$	\\
86	&	00	:	41	:	45.8	&	+	40	:	43	:	04	&		&	faint	&							&							&	0.456	&	$\mathrm{			<SNR>^1	}$	\\
87	&	00	:	41	:	48.3	&	+	41	:	07	:	06	&		&	faint	&							&							&	0.525	&	$\mathrm{	<hard>	174		}$	\\
88	&	00	:	41	:	49.6	&	+	41	:	01	:	07	&	mos	&	pl	&	0.4	$_{-	0.4	}^{+	1.6	}$	&	1.6	$_{-	0.9	}^{+	2.0	}$	&	1.16 (0.73)	&	$\mathrm{	<hard>	175		}$	\\
89	&	00	:	41	:	51.8	&	+	40	:	54	:	28	&	pn \& mos	&	nsa	&	f						&	5.22	$_{-	0.06	}^{+	0.04	}$	&	5.48 (4.14) / 8.00 (6.04)	&	$\mathrm{	<hard>	185		}$	\\
90	&	00	:	41	:	52.9	&	+	40	:	47	:	09	&	mos	&	pl	&	0.5	$_{-	0.5	}^{+	0.8	}$	&	1.8	$_{-	1.1	}^{+	2.1	}$	&	1.69 (0..99)	&	$\mathrm{	GlC	187		}$	\\
91	&	00	:	41	:	53.4	&	+	40	:	53	:	21	&	pn \& mos	&	pl	&	0.14	$_{-	0.04	}^{+	0.04	}$	&	2.1	$_{-	0.2	}^{+	0.2	}$	&	14.2 (1.71) / 15.5 (1.49)	&	$\mathrm{	<hard>	190		}$	\\
92	&	00	:	41	:	53.9	&	+	41	:	07	:	24	&	pn \& mos	&	pl	&	f						&	2.2	$_{-	0.2	}^{+	0.3	}$	&	6.98 (1.38) / 6.31 (1.17)	&	$\mathrm{	<SSS>	191		}$	\\
93	&	00	:	41	:	56.5	&	+	40	:	47	:	13	&	pn \& mos	&	pl	&	0.4	$_{-	0.1	}^{+	0.1	}$	&	1.7	$_{-	0.2	}^{+	0.3	}$	&	15.5 (3.62) / 16.1 (2.02)	&	$\mathrm{	<hard>	194		}$	\\
94	&	00	:	42	:	02.9	&	+	40	:	46	:	06	&	pn \& mos	&	pl	&	0.1	$_{-	0.1	}^{+	0.1	}$	&	2.0	$_{-	0.6	}^{+	0.6	}$	&	7.41 (2.51) / 9.63 (2.05)	&	$\mathrm{	<hard>	199		}$	\\
95	&	00	:	42	:	05.9	&	+	41	:	02	:	48	&	mos	&	bb	&	0.5	$_{-	0.5	}^{+	2.9	}$	&	0.9	$_{-	0.3	}^{+	0.5	}$	&	1.53 (1.47)	&	$\mathrm{	<GlC>	207		}$	\\
96	&	00	:	42	:	06.8	&	+	41	:	00	:	16	&	pn	&	pl	&	f						&	0.8	$_{-	0.4	}^{+	0.5	}$	&	9.19 (4.35)	&	$\mathrm{	<GlC>	208		}$	\\
97	&	00	:	42	:	07.6	&	+	41	:	04	:	36	&	pn \& mos	&	pl	&	0.17	$_{-	0.13	}^{+	0.12	}$	&	2.0	$_{-	0.3	}^{+	0.5	}$	&	2.44 (0.66) / 3.00 (0.69)	&	$\mathrm{	<hard>	213		}$	\\
98	&	00	:	42	:	09.8	&	+	40	:	50	:	38	&	pn \& mos	&	pl	&	0.48	$_{-	0.03	}^{+	0.04	}$	&	1.46	$_{-	0.06	}^{+	0.06	}$	&	288 (17) / 284 (11)	&	$\mathrm{	<hard>	222		}$	\\
99	&	00	:	42	:	10.8	&	+	41	:	06	:	46	&	pn	&	bb	&	f						&	1.8	$_{-	0.9	}^{+	3.7	}$	&	1.76 (1.49)	&	$\mathrm{	<hard>	225		}$	\\
100	&	00	:	42	:	11.8	&	+	40	:	53	:	37	&		&	faint	&							&							&	0.010	&	$\mathrm{	<hard>	229		}$	\\
101	&	00	:	42	:	15.6	&	+	41	:	01	:	14	&	pn \& mos	&	pl	&	0.27	$_{-	0.04	}^{+	0.04	}$	&	2.44	$_{-	0.14	}^{+	0.13	}$	&	28.6 (2.55) / 30.5 (2.39)	&	$\mathrm{	GlC	239		}$	\\
102	&	00	:	42	:	15.8	&	+	40	:	59	:	59	&	mos	&	pl	&	1.6	$_{-	1.6	}^{+	3.5	}$	&	3	$_{-	2	}^{+	4	}$	&	3.48 (3.10)	&	$\mathrm{			<hard>^1	}$	\\
103	&	00	:	42	:	16.3	&	+	40	:	48	:	15	&	pn \& mos	&	pl	&	0.4	$_{-	0.3	}^{+	0.6	}$	&	1.1	$_{-	0.4	}^{+	0.6	}$	&	4.53 (1.55) / 5.09 (1.58)	&	$\mathrm{	<SNR>	242		}$	\\
104	&	00	:	42	:	16.4	&	+	40	:	55	:	52	&	pn \& mos	&	pl	&	0.13	$_{-	0.03	}^{+	0.03	}$	&	1.75	$_{-	0.11	}^{+	0.11	}$	&	33.5 (2.71) / 31.6 (2.25)	&	$\mathrm{	<hard>	241		}$	\\
105	&	00	:	42	:	16.5	&	+	40	:	52	:	41	&	pn	&	pl	&	f						&	3	$_{-	2	}^{+	4	}$	&	2.19 (1.67)	&	$\mathrm{	<hard>	243		}$	\\
106	&	00	:	42	:	18.9	&	+	41	:	00	:	23	&	pn \& mos	&	br	&	0.5	$_{-	0.2	}^{+	0.5	}$	&	1.5	$_{-	0.9	}^{+	1.4	}$	&	2.54 (1.12) / 1.92 (0.80)	&	$\mathrm{			<hard>^1	}$	\\
107	&	00	:	42	:	22.0	&	+	40	:	59	:	23	&	pn	&	pl	&	0.19	$_{-	0.03	}^{+	0.04	}$	&	2.5	$_{-	0.2	}^{+	0.2	}$	&	48.8 (3.87)	&	$\mathrm{	<hard>	254		}$	\\
108	&	00	:	42	:	23.2	&	+	41	:	07	:	35	&	pn	&	pl	&	f						&	1.5	$_{-	0.3	}^{+	0.3	}$	&	5.18 (1.93)	&	$\mathrm{	<hard>	258		}$	\\
109	&	00	:	42	:	24.7	&	+	40	:	57	:	20	&	pn \& mos	&	br	&	f						&	5.6	$_{-	1.2	}^{+	1.8	}$	&	19.2 (2.7) / 20.8 (2.5)	&	$\mathrm{	GlC	261		}$	\\
110	&	00	:	42	:	26.0	&	+	40	:	54	:	52	&	pn	&	pl	&	0.21	$_{-	0.07	}^{+	0.07	}$	&	2.3	$_{-	0.3	}^{+	0.3	}$	&	18.0 (2.42)	&	$\mathrm{	<hard>	265		}$	\\
111	&	00	:	42	:	28.7	&	+	41	:	04	:	36	&	mos	&	bb	&	f						&	0.9	$_{-	0.3	}^{+	0.6	}$	&	2.24 (2.24)	&	$\mathrm{	<hard>	271		}$	\\
112	&	00	:	42	:	32.9	&	+	41	:	03	:	28	&	pn	&	pl	&	f						&	1.78	$_{-	0.05	}^{+	0.06	}$	&	95.7 (6.00)	&	$\mathrm{	GlC	282		}$	\\
113	&	00	:	42	:	34.7	&	+	40	:	57	:	19	&	pn \& mos	&	pl	&	0.7	$_{-	0.2	}^{+	0.7	}$	&	2.2	$_{-	0.4	}^{+	1.3	}$	&	2.98 (1.28) / 4.05 (1.67)	&	$\mathrm{	<hard>	288		}$	\\
114	&	00	:	42	:	34.9	&	+	40	:	48	:	40	&	pn \& mos	&	pl	&	3	$_{-	2	}^{+	3	}$	&	2.0	$_{-	0.4	}^{+	1.3	}$	&	5.42 (2.41) / 5.32 (2.57)	&	$\mathrm{	<hard>	289		}$	\\
115	&	00	:	42	:	36.2	&	+	40	:	58	:	48	&	pn \& mos	&	pl	&	0.20	$_{-	0.05	}^{+	0.16	}$	&	1.5	$_{-	0.2	}^{+	0.3	}$	&	4.91 (1.14) / 4.26 (0.68)	&	$\mathrm{	<hard>	293		}$	\\
116	&	00	:	42	:	42.3	&	+	40	:	51	:	50	&		&	faint	&							&							&	0.298	&	$\mathrm{	Gal	315		}$	\\
117	&	00	:	42	:	51.2	&	+	41	:	32	:	12	&	mos	&	pl	&	0.5	$_{-	0.5	}^{+	0.6	}$	&	1.9	$_{-	0.8	}^{+	1.1	}$	&	4.13 (1.75)	&	$\mathrm{			<hard>^1	}$	\\
118	&	00	:	42	:	51.8	&	+	41	:	31	:	10	&	mos	&	br	&	0.23	$_{-	0.01	}^{+	0.01	}$	&	11	$_{-	1	}^{+	2	}$	&	411 (14)	&	$\mathrm{	GlC	351		}$	\\
119	&	00	:	42	:	53.4	&	+	41	:	29	:	54	&	mos	&	pl	&	0.13	$_{-	0.13	}^{+	0.25	}$	&	1.4	$_{-	0.4	}^{+	0.4	}$	&	6.48 (1.63)	&	$\mathrm{	<hard>	355		}$	\\
120	&	00	:	42	:	57.7	&	+	41	:	39	:	13	&		&	faint	&							&							&	0.269	&	$\mathrm{	<hard>	363		}$	\\
121	&	00	:	42	:	58.8	&	+	41	:	37	:	33	&	pn	&	bb	&	0.4	$_{-	0.4	}^{+	1.1	}$	&	0.5	$_{-	0.2	}^{+	0.2	}$	&	1.07 (0.76)	&	$\mathrm{	<hard>	367		}$	\\
122	&	00	:	42	:	59.4	&	+	41	:	29	:	45	&	mos	&	pl	&	0.4	$_{-	0.2	}^{+	0.4	}$	&	1.9	$_{-	0.4	}^{+	0.7	}$	&	5.21 (1.54)	&	$\mathrm{	<hard>	370		}$	\\
123	&	00	:	43	:	01.4	&	+	41	:	30	:	18	&	mos	&	pl	&	0.18	$_{-	0.06	}^{+	0.07	}$	&	0.87	$_{-	0.06	}^{+	0.10	}$	&	73.5 (6.00)	&	$\mathrm{	GlC	377		}$	\\
124	&	00	:	43	:	03.8	&	+	41	:	38	:	46	&	pn \& mos	&	pl	&	0.18	$_{-	0.16	}^{+	0.22	}$	&	1.9	$_{-	0.6	}^{+	0.4	}$	&	1.83 (0.77) / 2.09 (0.69)	&	$\mathrm{	<AGN>	387		}$	\\
125	&	00	:	43	:	05.2	&	+	41	:	40	:	24	&		&	faint	&							&							&	0.222	&	$\mathrm{	<fg star>	389		}$	\\
126	&	00	:	43	:	06.7	&	+	41	:	35	:	23	&	pn	&	pl	&	f						&	2.5	$_{-	0.9	}^{+	1.2	}$	&	0.88 (0.69)	&	$\mathrm{	<hard>	393		}$	\\
127	&	00	:	43	:	08.2	&	+	41	:	46	:	03	&	pn	&	pl	&	0.5	$_{-	0.2	}^{+	0.8	}$	&	1.8	$_{-	0.8	}^{+	1.2	}$	&	2.83 (1.15)	&	$\mathrm{	<hard>	400		}$	\\
128	&	00	:	43	:	08.6	&	+	41	:	32	:	13	&	pn	&	br	&	f						&	0.4						&	0.95 (0.9)	&	$\mathrm{	<hard>	402		}$	\\
129	&	00	:	43	:	10.4	&	+	41	:	38	:	54	&	pn	&	pl	&	5	$_{-	5	}^{+	46	}$	&	1.9	$_{-	2.2	}^{+	8.1	}$	&	2.05 (1.91)	&	$\mathrm{	<hard>	407		}$	\\
130	&	00	:	43	:	17.9	&	+	41	:	39	:	16	&	pn \& mos	&	pl	&	0.4	$_{-	0.3	}^{+	0.6	}$	&	1.5	$_{-	0.7	}^{+	0.9	}$	&	2.09 (0.93) / 2.33 (0.98)	&	$\mathrm{	<hard>	425		}$	\\
131	&	00	:	43	:	18.6	&	+	41	:	43	:	14	&	pn	&	br	&	0.23	$_{-	0.15	}^{+	0.47	}$	&	1.1	$_{-	0.6	}^{+	1.0	}$	&	0.68 (0.67)	&	$\mathrm{	<fg star>	429		}$	\\
132	&	00	:	43	:	23.4	&	+	41	:	31	:	47	&	pn \& mos	&	pl	&	f						&	1.8	$_{-	0.4	}^{+	0.4	}$	&	1.35 (0.74) / 1.51 (0.69)	&	$\mathrm{	<hard>	441		}$	\\
133	&	00	:	43	:	24.9	&	+	41	:	35	:	56	&	pn \& mos	&	pl	&	f						&	1.5	$_{-	0.4	}^{+	0.4	}$	&	1.41 (0.56) / 1.12 (0.49)	&	$\mathrm{	<hard>	444		}$	\\
134	&	00	:	43	:	25.4	&	+	41	:	36	:	52	&	pn \& mos	&	pl	&	0.38	$_{-	0.18	}^{+	0.29	}$	&	2.3	$_{-	0.5	}^{+	0.4	}$	&	1.37 (0.49) / 1.25 (0.41)	&	$\mathrm{	<hard>	445		}$	\\
135	&	00	:	43	:	31.2	&	+	41	:	40	:	49	&	pn \& mos	&	pl	&	f						&	1.9	$_{-	0.7	}^{+	1.0	}$	&	0.56 (0.44) / 0.94 (0.69)	&	$\mathrm{	<hard>	460		}$	\\
136	&	00	:	43	:	31.6	&	+	41	:	45	:	50	&	pn \& mos	&	pl	&	0.8	$_{-	0.8	}^{+	2.3	}$	&	1.1	$_{-	1.1	}^{+	2.1	}$	&	1.96 (1.41) / 1.94 (1.58)	&	$\mathrm{	<hard>	462		}$	\\
137	&	00	:	43	:	35.7	&	+	41	:	33	:	22	&	pn \& mos	&	pl	&	0.5	$_{-	0.3	}^{+	0.7	}$	&	1.3	$_{-	0.5	}^{+	0.8	}$	&	1.23 (0.51) / 1.41 (0.56)	&	$\mathrm{	<hard>	468		}$	\\
138	&	00	:	43	:	39.0	&	+	41	:	26	:	54	&	pn \& mos	&	bb	&	f						&	1.33	$_{-	0.01	}^{+	0.01	}$	&	3.74 (.081) / 3.89 (0.75)	&	$\mathrm{	SNR	475		}$	\\
139	&	00	:	43	:	40.5	&	+	41	:	41	:	05	&	pn \& mos	&	pl	&	1.7	$_{-	1.1	}^{+	3.8	}$	&	2.4	$_{-	1.0	}^{+	2.7	}$	&	1.57 (1.11) / 1.81 (1.25)	&	$\mathrm{	<AGN>	477		}$	\\
140	&	00	:	43	:	41.5	&	+	41	:	42	:	26	&	pn	&	bb	&	f						&	0.16	$_{-	0.02	}^{+	0.03	}$	&	0.62 (0.30)	&	$\mathrm{	<fg star>	479		}$	\\
141	&	00	:	43	:	42.7	&	+	41	:	33	:	11	&		&	13060	&							&							&	0.422	&	$\mathrm{	<hard>	482		}$	\\
142	&	00	:	43	:	43.9	&	+	41	:	28	:	47	&	pn \& mos	&	pl	&	0.30	$_{-	0.08	}^{+	0.17	}$	&	1.7	$_{-	0.2	}^{+	0.3	}$	&	7.14 (1.09) / 4.66 (0.86)	&	$\mathrm{	<GlC>	483		}$	\\
143	&	00	:	43	:	45.5	&	+	41	:	36	:	57	&	pn \& mos	&	pl	&	f						&	1.84	$_{-	0.15	}^{+	0.16	}$	&	4.16 (1.31) / 5.08 (1.00)	&	$\mathrm{	GlC	488		}$	\\
144	&	00	:	43	:	45.5	&	+	41	:	27	:	09	&	pn	&	pl	&	0.5	$_{-	0.5	}^{+	0.9	}$	&	2.0	$_{-	1.1	}^{+	2.5	}$	&	1.12 (0.59)	&	$\mathrm{	<hard>	489		}$	\\
145	&	00	:	43	:	46.8	&	+	41	:	38	:	40	&	mos	&	br	&	f						&	0.4	$_{-	0.2	}^{+	0.5	}$	&	0.62 (0.54)	&	$\mathrm{	<fg star>	492		}$	\\
146	&	00	:	43	:	47.1	&	+	41	:	27	:	47	&	pn \& mos	&	bb	&	0.20	$_{-	0.06	}^{+	0.15	}$	&	0.17	$_{-	0.03	}^{+	0.03	}$	&	2.86 (0.58) / 3.07 (0.58)	&	$\mathrm{	<fg star>	493		}$	\\
147	&	00	:	43	:	47.2	&	+	41	:	33	:	20	&	pn \& mos	&	pl	&	0.5	$_{-	0.4	}^{+	0.8	}$	&	1.5	$_{-	0.5	}^{+	0.8	}$	&	0.95 (0.40) / 1.15 (0.47)	&	$\mathrm{	<hard>	494		}$	\\
148	&	00	:	43	:	48.1	&	+	41	:	35	:	34	&		&	faint	&							&							&	0.304	&	$\mathrm{	<hard>	496		}$	\\
149	&	00	:	43	:	53.9	&	+	41	:	31	:	05	&	mos	&	pl	&	f						&	0.9	$_{-	0.7	}^{+	0.7	}$	&	1.01 (0.73)	&	$\mathrm{	<hard>	505		}$	\\
150	&	00	:	43	:	55.2	&	+	41	:	32	:	54	&	pn	&	pl	&	0.3	$_{-	0.3	}^{+	4.6	}$	&	3	$_{-	6	}^{+	7	}$	&	1.03 (0.95)	&	$\mathrm{	<hard>	506		}$	\\
151	&	00	:	43	:	56.1	&	+	41	:	22	:	04	&	mos	&	pl	&	0.20	$_{-	0.17	}^{+	0.16	}$	&	2.1	$_{-	0.5	}^{+	0.8	}$	&	4.13 (1.06)	&	$\mathrm{	GlC	508		}$	\\
152	&	00	:	43	:	56.6	&	+	41	:	49	:	40	&		&	faint	&							&							&	-	&	$\mathrm{			<hard>^1	}$	\\
153	&	00	:	43	:	57.5	&	+	41	:	43	:	48	&	mos	&	pl	&	0.5	$_{-	0.3	}^{+	0.9	}$	&	2.1	$_{-	0.9	}^{+	1.3	}$	&	1.93 (0.88)	&	$\mathrm{	<fg star>	513		}$	\\
154	&	00	:	43	:	57.5	&	+	41	:	30	:	57	&	pn \& mos	&	pl	&	f						&	1.63	$_{-	0.11	}^{+	0.12	}$	&	4.68 (0.75) / 4.72 (0.60)	&	$\mathrm{	<hard>	512		}$	\\
155	&	00	:	44	:	00.5	&	+	41	:	28	:	03	&	mos	&	pl	&	f						&	1.3	$_{-	0.6	}^{+	0.7	}$	&	1.63 (1.12)	&	$\mathrm{	<hard>	515		}$	\\
156	&	00	:	44	:	01.8	&	+	41	:	40	:	30	&	pn	&	pl	&	0.6	$_{-	0.4	}^{+	0.9	}$	&	3.1	$_{-	1.1	}^{+	2.3	}$	&	1.63 (1.00)	&	$\mathrm{	<hard>	517		}$	\\
157	&	00	:	44	:	02.7	&	+	41	:	39	:	28	&	pn \& mos	&	pl	&	0.9	$_{-	0.2	}^{+	0.3	}$	&	3.1	$_{-	0.4	}^{+	0.5	}$	&	5.90 (2.36) / 5.26 (1.69)	&	$\mathrm{	<hard>	519		}$	\\
158	&	00	:	44	:	04.0	&	+	41	:	44	:	24	&	pn	&	pl	&	0.6	$_{-	0.3	}^{+	0.8	}$	&	2.4	$_{-	1.0	}^{+	1.6	}$	&	1.66 (0.85)	&	$\mathrm{	<hard>	521		}$	\\
159	&	00	:	44	:	04.9	&	+	41	:	21	:	28	&	mos	&	br	&	7	$_{-	5	}^{+	8	}$	&	0.7	$_{-	0.7	}^{+	1.9	}$	&	58.3 (58.2)	&	$\mathrm{	<AGN>	524		}$	\\
160	&	00	:	44	:	06.5	&	+	41	:	38	:	57	&	mos	&	pl	&	f						&	1.0	$_{-	0.7	}^{+	0.6	}$	&	1.41 (0.73)	&	$\mathrm{	<hard>	525		}$	\\
161	&	00	:	44	:	07.8	&	+	41	:	56	:	07	&		&	faint	&							&							&	1.020	&	$\mathrm{	<hard>	528		}$	\\
162	&	00	:	44	:	10.1	&	+	41	:	33	:	45	&	pn \& mos	&	pl	&	f						&	3.7	$_{-	0.9	}^{+	1.0	}$	&	0.73 (0.51) / 0.56 (0.39)	&	$\mathrm{	<hard>	532		}$	\\
163	&	00	:	44	:	12.0	&	+	41	:	31	:	50	&	pn \& mos	&	pl	&	0.13	$_{-	0.07	}^{+	0.05	}$	&	2.0	$_{-	0.2	}^{+	0.2	}$	&	2.84 (0.52) / 2.78 (0.44)	&	$\mathrm{	<hard>	535		}$	\\
164	&	00	:	44	:	12.1	&	+	41	:	45	:	13	&	pn \& mos	&	pl	&	f						&	1.9	$_{-	0.6	}^{+	0.6	}$	&	1.50 (0.83) / 1.32 (0.65)	&	$\mathrm{	<hard>	536		}$	\\
165	&	00	:	44	:	13.2	&	+	41	:	56	:	52	&		&	faint	&							&							&	1.528	&	$\mathrm{	<hard>	539		}$	\\
166	&	00	:	44	:	15.9	&	+	41	:	30	:	59	&	pn \& mos	&	pl	&	0.9	$_{-	0.3	}^{+	0.5	}$	&	1.3	$_{-	0.2	}^{+	0.3	}$	&	11.8 (1.7) / 11.7 (2.0)	&	$\mathrm{	XRB	544		}$	\\
167	&	00	:	44	:	16.5	&	+	41	:	26	:	29	&		&	faint	&							&							&	0.207	&	$\mathrm{	<fg star>	545		}$	\\
168	&	00	:	44	:	17.9	&	+	41	:	50	:	24	&		&	faint	&							&							&	0.575	&	$\mathrm{	<fg star>	546		}$	\\
169	&	00	:	44	:	18.3	&	+	41	:	51	:	33	&		&	faint	&							&							&	1.202	&	$\mathrm{	<hard>	547		}$	\\
170	&	00	:	44	:	18.9	&	+	41	:	32	:	11	&	mos	&	pl	&	1.4	$_{-	1.3	}^{+	6.8	}$	&	2.8	$_{-	1.5	}^{+	4.4	}$	&	1.79 (1.34)	&	$\mathrm{	<hard>	548		}$	\\
171	&	00	:	44	:	20.0	&	+	41	:	34	:	07	&	pn \& mos	&	pl	&	f						&	2.2	$_{-	0.4	}^{+	0.4	}$	&	0.50 (0.23) / 0.44 (0.18)	&	$\mathrm{	<hard>	549		}$	\\
172	&	00	:	44	:	20.7	&	+	41	:	35	:	43	&	pn \& mos	&	pl	&	f						&	1.1	$_{-	0.6	}^{+	0.6	}$	&	1.26 (0.91) / 1.09 (0.73)	&	$\mathrm{	<hard>	550		}$	\\
173	&	00	:	44	:	22.9	&	+	41	:	45	:	07	&	pn \& mos	&	pl	&	f						&	1.93	$_{-	0.12	}^{+	0.13	}$	&	7.59 (1.20) / 13.9 (1.43)	&	$\mathrm{	<hard>	551		}$	\\
174	&	00	:	44	:	23.7	&	+	42	:	00	:	08	&	pn	&	bb	&	f						&	0.16	$_{-	0.03	}^{+	0.04	}$	&	1.17 (0.82)	&	$\mathrm{	<fg star>	553		}$	\\
175	&	00	:	44	:	24.7	&	+	41	:	32	:	01	&	pn \& mos	&	pl	&	0.20	$_{-	0.06	}^{+	0.04	}$	&	1.92	$_{-	0.15	}^{+	0.10	}$	&	5.92 (0.79) / 6.36 (0.65)	&	$\mathrm{	<hard>	555		}$	\\
176	&	00	:	44	:	25.5	&	+	41	:	36	:	35	&	pn \& mos	&	bb	&	0.24	$_{-	0.09	}^{+	0.11	}$	&	0.16	$_{-	0.02	}^{+	0.02	}$	&	4.84 (0.71) / 4.58 (0.79)	&	$\mathrm{	<fg star>	556		}$	\\
177	&	00	:	44	:	25.8	&	+	41	:	30	:	35	&	pn \& mos	&	pl	&	0.24	$_{-	0.13	}^{+	0.20	}$	&	2.6	$_{-	0.4	}^{+	1.0	}$	&	1.95 (0.60) / 2.02 (0.62)	&	$\mathrm{	<hard>	558		}$	\\
178	&	00	:	44	:	28.0	&	+	41	:	42	:	09	&	pn	&	br	&	f						&	0.4	$_{-	0.4	}^{+	0.5	}$	&	4.71 (1.03)	&	$\mathrm{	<hard>	561		}$	\\
179	&	00	:	44	:	29.5	&	+	41	:	54	:	49	&		&	faint	&							&							&	-	&	$\mathrm{			<hard>^1	}$	\\
180	&	00	:	44	:	30.4	&	+	41	:	40	:	40	&		&	faint	&							&							&	0.399	&	$\mathrm{	<hard>	563		}$	\\
181	&	00	:	44	:	30.6	&	+	41	:	23	:	06	&	mos	&	bb	&	f						&	0.7	$_{-	0.2	}^{+	0.3	}$	&	1.67 (1.06)	&	$\mathrm{	<hard>	564		}$	\\
182	&	00	:	44	:	32.2	&	+	41	:	25	:	23	&	pn	&	pl	&	f						&	2.5	$_{-	1.0	}^{+	1.2	}$	&	1.18 (0.76)	&	$\mathrm{	fg Star	565		}$	\\
183	&	00	:	44	:	33.5	&	+	42	:	06	:	07	&		&	faint	&							&							&	1.240	&	$\mathrm{	<hard>	567		}$	\\
184	&	00	:	44	:	36.4	&	+	41	:	25	:	31	&		&	faint	&							&							&	0.134	&	$\mathrm{			<hard>^1	}$	\\
185	&	00	:	44	:	36.8	&	+	42	:	04	:	36	&	pn	&	pl	&	0.3	$_{-	0.3	}^{+	1.1	}$	&	1.1	$_{-	0.8	}^{+	1.3	}$	&	4.91 (2.46)	&	$\mathrm{	<hard>	569		}$	\\
186	&	00	:	44	:	37.8	&	+	41	:	45	:	15	&	pn \& mos	&	pl	&	0.16	$_{-	0.07	}^{+	0.06	}$	&	2.2	$_{-	0.2	}^{+	0.3	}$	&	7.69 (1.21) / 9.02 (1.19)	&	$\mathrm{	<AGN>	570		}$	\\
187	&	00	:	44	:	37.9	&	+	41	:	45	:	14	&	mos	&	pl	&	0.12	$_{-	0.12	}^{+	0.21	}$	&	2.2	$_{-	0.6	}^{+	0.7	}$	&	5.77 (1.73)	&	$\mathrm{	<AGN>	570		}$	\\
188	&	00	:	44	:	38.0	&	+	41	:	40	:	09	&	pn \& mos	&	pl	&	f						&	8	$_{-	8	}^{+	2	}$	&	44.3 (43.5) / 62.4 (61.5)	&	$\mathrm{			<hard>^1	}$	\\
189	&	00	:	44	:	38.7	&	+	41	:	31	:	47	&	pn \& mos	&	pl	&	0.6	$_{-	0.3	}^{+	0.2	}$	&	2.2	$_{-	1.2	}^{+	2.9	}$	&	1.11 (0.76) / 1.55 (1.16)	&	$\mathrm{	<hard>	573		}$	\\
190	&	00	:	44	:	42.7	&	+	41	:	53	:	41	&	pn \& mos	&	pl	&	6	$_{-	4	}^{+	9	}$	&	0.4	$_{-	0.9	}^{+	1.4	}$	&	8.43 (5.94) / 14.8 (9.65)	&	$\mathrm{	<hard>	577		}$	\\
191	&	00	:	44	:	43.3	&	+	41	:	26	:	28	&	pn	&	pl	&	f						&	3.7	$_{-	1.1	}^{+	1.6	}$	&	1.23 (0.93)	&	$\mathrm{			<hard>^1	}$	\\
192	&	00	:	44	:	43.6	&	+	41	:	46	:	47	&		&	faint	&							&							&	1.084	&	$\mathrm{	<AGN>	?		}$	\\
193	&	00	:	44	:	44.8	&	+	41	:	51	:	55	&	pn \& mos	&	pl	&	f						&	1.6	$_{-	0.4	}^{+	0.5	}$	&	3.12 (1.20) / 3.10 (1.16)	&	$\mathrm{	<hard>	581		}$	\\
194	&	00	:	44	:	45.1	&	+	41	:	46	:	45	&		&	faint	&							&							&	0.156	&	$\mathrm{	<AGN>	580		}$	\\
195	&	00	:	44	:	46.0	&	+	41	:	42	:	22	&	pn \& mos	&	bb	&	f						&	1.4	$_{-	0.5	}^{+	1.4	}$	&	1.62 (1.47) / 2.93 (2.50)	&	$\mathrm{	<hard>	582		}$	\\
196	&	00	:	44	:	47.2	&	+	41	:	29	:	21	&		&	faint	&							&							&	0.260	&	$\mathrm{	SNR	583		}$	\\
197	&	00	:	44	:	47.3	&	+	41	:	44	:	14	&	pn \& mos	&	pl	&	f						&	1.0	$_{-	0.7	}^{+	0.8	}$	&	4.55 (3.98) / 2.88 (2.04)	&	$\mathrm{	<hard>	584		}$	\\
198	&	00	:	44	:	48.8	&	+	41	:	58	:	13	&		&	faint	&							&							&	1.386	&	$\mathrm{	<hard>	586		}$	\\
199	&	00	:	44	:	49.3	&	+	41	:	47	:	27	&		&	faint	&							&							&	1.465	&	$\mathrm{	<hard>	588		}$	\\
200	&	00	:	44	:	49.6	&	+	41	:	47	:	28	&	pn	&	pl	&	f						&	2						&	1.47 (0.97)	&	$\mathrm{	<hard>	588		}$	\\
201	&	00	:	44	:	50.9	&	+	41	:	29	:	06	&	pn \& mos	&	pl	&	f						&	3.0	$_{-	0.3	}^{+	0.3	}$	&	2.15 (0.74) / 2.14 (0.59)	&	$\mathrm{	SNR	589		}$	\\
202	&	00	:	44	:	51.3	&	+	41	:	27	:	11	&	pn \& mos	&	pl	&	f						&	1.9	$_{-	0.2	}^{+	0.2	}$	&	4.18 (1.20) / 8.82 (1.98)	&	$\mathrm{			<hard>^1	}$	\\
203	&	00	:	44	:	51.5	&	+	41	:	38	:	33	&	pn	&	pl	&	f						&	2.1	$_{-	1.9	}^{+	6.4	}$	&	2.76 (2.61)	&	$\mathrm{	<fg star>	590		}$	\\
204	&	00	:	44	:	53.2	&	+	42	:	02	:	15	&		&	faint	&							&							&	0.178	&	$\mathrm{	<fg star>	593		}$	\\
205	&	00	:	44	:	55.3	&	+	41	:	34	:	41	&	pn \& mos	&	pl	&	0.5	$_{-	0.1	}^{+	0.2	}$	&	1.8	$_{-	0.2	}^{+	0.2	}$	&	10.3 (1.9) / 10.7 (1.6)	&	$\mathrm{	<AGN>	595		}$	\\
206	&	00	:	44	:	56.4	&	+	41	:	59	:	37	&	pn \& mos	&	br	&	0.17	$_{-	0.07	}^{+	0.12	}$	&	0.49	$_{-	0.16	}^{+	0.18	}$	&	4.02 (0.85) / 3.37 (0.95)	&	$\mathrm{	<fg star>	598		}$	\\
207	&	00	:	44	:	58.4	&	+	41	:	46	:	23	&	pn	&	pl	&	f						&	2.0	$_{-	1.2	}^{+	1.8	}$	&	2.77 (2.55)	&	$\mathrm{	<hard>	602		}$	\\
208	&	00	:	44	:	58.5	&	+	41	:	46	:	21	&		&	faint	&							&							&	1.029	&	$\mathrm{	<hard>	602		}$	\\
209	&	00	:	44	:	59.2	&	+	41	:	40	:	06	&	pn	&	bb	&	f						&	0.8	$_{-	0.8	}^{+	199	}$	&	0.66 (0.6)	&	$\mathrm{	<hard>	603		}$	\\
210	&	00	:	45	:	00.5	&	+	41	:	27	:	04	&	pn	&	pl	&	0.5	$_{-	0.4	}^{+	1.1	}$	&	2.7	$_{-	0.7	}^{+	2.2	}$	&	2.82 (1.34)	&	$\mathrm{	<hard>	606		}$	\\
211	&	00	:	45	:	01.2	&	+	41	:	56	:	09	&		&	faint	&							&							&	1.036	&	$\mathrm{	<AGN>	607		}$	\\
212	&	00	:	45	:	06.3	&	+	42	:	06	:	19	&		&	faint	&							&							&	4.461	&	$\mathrm{	<hard>	612		}$	\\
213	&	00	:	45	:	06.9	&	+	42	:	03	:	00	&	pn	&	pl	&	f						&	0.6	$_{-	0.5	}^{+	0.6	}$	&	2.23 (1.67)	&	$\mathrm{	<AGN>	614		}$	\\
214	&	00	:	45	:	07.4	&	+	41	:	53	:	56	&		&	faint	&							&							&	1.432	&	$\mathrm{	<fg star>	615		}$	\\
215	&	00	:	45	:	09.8	&	+	42	:	02	:	38	&	mos	&	br	&	f						&	0.36	$_{-	0.13	}^{+	0.15	}$	&	2.86 (2.26)	&	$\mathrm{	<fg star>	616		}$	\\
216	&	00	:	45	:	11.6	&	+	41	:	45	:	59	&	pn \& mos	&	pl	&	0.15	$_{-	0.15	}^{+	0.28	}$	&	1.6	$_{-	0.6	}^{+	0.5	}$	&	3.77 (1.14) / 2.38 (0.89)	&	$\mathrm{	<hard>	617		}$	\\
217	&	00	:	45	:	13.6	&	+	41	:	35	:	30	&		&	faint	&							&							&	0.256	&	$\mathrm{			<hard>^1	}$	\\
218	&	00	:	45	:	13.8	&	+	41	:	36	:	17	&	pn \& mos	&	br	&	0.38	$_{-	0.14	}^{+	0.08	}$	&	0.24	$_{-	0.04	}^{+	0.06	}$	&	39.6 (38.4) / 52.3 (51.3)	&	$\mathrm{	SNR	621		}$	\\
219	&	00	:	45	:	15.1	&	+	41	:	50	:	36	&	pn	&	bb	&	f						&	0.22	$_{-	0.06	}^{+	0.09	}$	&	0.62 (0.44)	&	$\mathrm{	<hard>	622		}$	\\
220	&	00	:	45	:	18.5	&	+	41	:	39	:	35	&	pn	&	pl	&	2	$_{-	2	}^{+	6	}$	&	1.5	$_{-	1.8	}^{+	3.6	}$	&	2.54 (2.35)	&	$\mathrm{	<hard>	626		}$	\\
221	&	00	:	45	:	19.7	&	+	42	:	09	:	08	&		&	faint	&							&							&	0.471	&	$\mathrm{	<fg star>	628		}$	\\
222	&	00	:	45	:	23.4	&	+	41	:	51	:	58	&	pn	&	pl	&	f						&	0.3	$_{-	0.8	}^{+	0.7	}$	&	2.47 (1.84)	&	$\mathrm{	<hard>	632		}$	\\
223	&	00	:	45	:	25.6	&	+	41	:	53	:	29	&		&	faint	&							&							&	1.014	&	$\mathrm{	<hard>	633		}$	\\
224	&	00	:	45	:	26.1	&	+	41	:	44	:	30	&		&	faint	&							&							&	0.769	&	$\mathrm{	<hard>	635		}$	\\
225	&	00	:	45	:	26.1	&	+	41	:	43	:	12	&		&	faint	&							&							&	0.925	&	$\mathrm{	<hard>	634		}$	\\
226	&	00	:	45	:	26.7	&	+	41	:	56	:	33	&	pn	&	pl	&	f						&	0.9	$_{-	0.9	}^{+	0.9	}$	&	1.66 (1.32)	&	$\mathrm{	<hard>	636		}$	\\
227	&	00	:	45	:	27.1	&	+	42	:	00	:	17	&	pn	&	pl	&	0.4	$_{-	0.4	}^{+	1.7	}$	&	2.0	$_{-	1.1	}^{+	5.6	}$	&	1.91 (1.15)	&	$\mathrm{	<hard>	637		}$	\\
228	&	00	:	45	:	28.3	&	+	41	:	46	:	05	&		&	faint	&							&							&	1.081	&	$\mathrm{	<SNR>	642		}$	\\
229	&	00	:	45	:	31.2	&	+	42	:	01	:	44	&	pn	&	bb	&	2	$_{-	2	}^{+	10	}$	&	0.6	$_{-	0.4	}^{+	0.6	}$	&	1.94 (1.93)	&	$\mathrm{	<hard>	645		}$	\\
230	&	00	:	45	:	31.3	&	+	42	:	12	:	48	&		&	faint	&							&							&	0.578	&	$\mathrm{	<hard>	647		}$	\\
231	&	00	:	45	:	32.5	&	+	41	:	55	:	07	&	mos	&	pl	&	1	$_{-	1	}^{+	14	}$	&	1.6	$_{-	1.8	}^{+	8.4	}$	&	1.65 (1.40)	&	$\mathrm{	<hard>	648		}$	\\
232	&	00	:	45	:	33.0	&	+	42	:	10	:	58	&	pn \& mos	&	pl	&	0.6	$_{-	0.4	}^{+	1.2	}$	&	4	$_{-	2	}^{+	6	}$	&	10.9 (8.3) / 8.40 (6.81)	&	$\mathrm{	<fg star>	649		}$	\\
233	&	00	:	45	:	33.5	&	+	42	:	08	:	07	&		&	faint	&							&							&	0.422	&	$\mathrm{	<hard>	650		}$	\\
234	&	00	:	45	:	34.7	&	+	42	:	17	:	49	&	mos	&	pl	&	f						&	0.9	$_{-	0.6	}^{+	0.7	}$	&	3.56 (2.45)	&	$\mathrm{	<hard>	652		}$	\\
235	&	00	:	45	:	35.5	&	+	42	:	20	:	32	&	pn \& mos	&	pl	&	0.17	$_{-	0.17	}^{+	0.57	}$	&	1.7	$_{-	0.9	}^{+	1.7	}$	&	1.72 (1.02) / 2.45 (1.38)	&	$\mathrm{	<hard>	654		}$	\\
236	&	00	:	45	:	38.0	&	+	42	:	12	:	33	&	mos	&	pl	&	f						&	0.9	$_{-	0.5	}^{+	0.6	}$	&	2.97 (1.95)	&	$\mathrm{	<hard>	661		}$	\\
237	&	00	:	45	:	38.8	&	+	41	:	56	:	16	&		&	faint	&							&		$_{-		}^{+		}$	&	1.235	&	$\mathrm{	<fg star>	662		}$	\\
238	&	00	:	45	:	40.3	&	+	42	:	08	:	05	&	pn \& mos	&	diskbb	&	0.13	$_{-	0.07	}^{+	0.07	}$	&	0.25	$_{-	0.05	}^{+	0.05	}$	&	37.1 (11.71) / 38.8 (14.7)	&	$\mathrm{	<fg star>	663		}$	\\
239	&	00	:	45	:	40.5	&	+	42	:	08	:	07	&	pn \& mos	&	bb	&	f						&	0.17	$_{-	0.01	}^{+	0.01	}$	&	11.6 (2.1) / 11.4 (1.9)	&	$\mathrm{	<fg star>	663		}$	\\
240	&	00	:	45	:	41.7	&	+	42	:	23	:	24	&	pn \& mos	&	pl	&	0.12	$_{-	0.12	}^{+	0.57	}$	&	2.6	$_{-	1.2	}^{+	3.7	}$	&	1.29 (0.71) / 1.09 (0.71)	&	$\mathrm{	<hard>	666		}$	\\
241	&	00	:	45	:	42.8	&	+	42	:	14	:	20	&	pn	&	pl	&	0.8	$_{-	0.6	}^{+	1.7	}$	&	3.7	$_{-	1.7	}^{+	6.1	}$	&	8.17 (7.41)	&	$\mathrm{	<hard>	670		}$	\\
242	&	00	:	45	:	43.8	&	+	42	:	08	:	42	&		&	faint	&							&							&	0.771	&	$\mathrm{			<hard>^1	}$	\\
243	&	00	:	45	:	44.1	&	+	42	:	08	:	44	&		&	faint	&							&							&	0.496	&	$\mathrm{			<hard>^1	}$	\\
244	&	00	:	45	:	44.8	&	+	41	:	58	:	58	&	pn	&	pl	&	0.33	$_{-	0.17	}^{+	0.19	}$	&	2.0	$_{-	0.4	}^{+	0.6	}$	&	5.48 (1.22)	&	$\mathrm{	<hard>	673		}$	\\
245	&	00	:	45	:	45.5	&	+	41	:	49	:	33	&		&	faint	&							&							&	0.870	&	$\mathrm{	<hard>	674		}$	\\
246	&	00	:	45	:	45.8	&	+	41	:	50	:	30	&		&	faint	&							&							&	1.236	&	$\mathrm{	<fg star>	675		}$	\\
247	&	00	:	45	:	50.9	&	+	41	:	58	:	34	&		&	faint	&							&							&	1.046	&	$\mathrm{			<hard>^1	}$	\\
248	&	00	:	45	:	51.5	&	+	42	:	04	:	20	&		&	faint	&							&							&	1.020	&	$\mathrm{	<hard>	682		}$	\\
249	&	00	:	45	:	53.4	&	+	42	:	16	:	10	&	pn	&	br	&	f						&	0.5	$_{-	0.2	}^{+	0.9	}$	&	0.63 (0.53)	&	$\mathrm{			<hard>^1	}$	\\
250	&	00	:	45	:	54.7	&	+	42	:	13	:	11	&		&	faint	&							&							&	0.109	&	$\mathrm{			<hard>^1	}$	\\
251	&	00	:	45	:	55.2	&	+	41	:	52	:	11	&		&	faint	&							&							&	1.238	&	$\mathrm{	<hard>	687		}$	\\
252	&	00	:	45	:	56.0	&	+	42	:	12	:	33	&	pn \& mos	&	pl	&	0.3	$_{-	0.3	}^{+	1.3	}$	&	2.6	$_{-	1.4	}^{+	0.7	}$	&	1.58 (0.79) / 1.52 (0.85)	&	$\mathrm{	<hard>	690		}$	\\
253	&	00	:	45	:	56.9	&	+	41	:	48	:	32	&		&	faint	&							&							&	0.793	&	$\mathrm{			<hard>^1	}$	\\
254	&	00	:	45	:	57.9	&	+	42	:	26	:	47	&	pn \& mos	&	pl	&	f						&	2.3	$_{-	0.3	}^{+	0.4	}$	&	5.31 (1.54) / 5.43 (1.11)	&	$\mathrm{	<hard>	694		}$	\\
255	&	00	:	45	:	58.2	&	+	42	:	02	:	59	&	pn	&	pl	&	f						&	2.0	$_{-	0.6	}^{+	0.7	}$	&	3.25 (1.63)	&	$\mathrm{	<fg star>	693		}$	\\
256	&	00	:	45	:	58.8	&	+	42	:	04	:	25	&	pn \& mos	&	pl	&	0.3	$_{-	0.2	}^{+	0.4	}$	&	1.2	$_{-	0.5	}^{+	0.4	}$	&	4.68 (1.64) / 6.04 (2.10)	&	$\mathrm{	<hard>	696		}$	\\
257	&	00	:	45	:	59.0	&	+	42	:	04	:	21	&		&	faint	&							&							&	0.602	&	$\mathrm{	<hard>	696		}$	\\
258	&	00	:	46	:	00.2	&	+	42	:	10	:	31	&	pn \& mos	&	pl	&	f						&	0.03	$_{-	0.8	}^{+	0.7	}$	&	4.22 (3.18) / 3.64 (2.63)	&	$\mathrm{	<hard>	698		}$	\\
259	&	00	:	46	:	02.9	&	+	42	:	24	:	31	&	pn \& mos	&	pl	&	0.3	$_{-	0.1	}^{+	0.3	}$	&	3.8	$_{-	0.9	}^{+	1.5	}$	&	3.20 (1.82) / 4.21 (2.18)	&	$\mathrm{	<fg star>	701		}$	\\
260	&	00	:	46	:	03.6	&	+	42	:	13	:	22	&		&	faint	&							&							&	0.284	&	$\mathrm{			<hard>^1	}$	\\
261	&	00	:	46	:	04.6	&	+	41	:	49	:	47	&		&	faint	&							&							&	0.982	&	$\mathrm{	<SNR>	704		}$	\\
262	&	00	:	46	:	05.1	&	+	41	:	51	:	44	&		&	faint	&							&							&	1.186	&	$\mathrm{	<hard>	705		}$	\\
263	&	00	:	46	:	05.1	&	+	42	:	25	:	52	&	mos	&	pl	&	5	$_{-	5	}^{+	12	}$	&	0.3	$_{-	1.5	}^{+	2.4	}$	&	6.69 (6.08)	&	$\mathrm{	<hard>	706		}$	\\
264	&	00	:	46	:	05.5	&	+	42	:	20	:	29	&	pn \& mos	&	pl	&	0.7	$_{-	0.6	}^{+	1.8	}$	&	2.0	$_{-	1.0	}^{+	3.2	}$	&	1.38 (0.92) / 1.82 (1.23)	&	$\mathrm{	<AGN>	707		}$	\\
265	&	00	:	46	:	08.2	&	+	42	:	10	:	53	&	mos	&	bb	&	0.3	$_{-	0.3	}^{+	1.5	}$	&	1.0	$_{-	0.3	}^{+	0.5	}$	&	1.50 (1.23)	&	$\mathrm{	<hard>	715		}$	\\
266	&	00	:	46	:	08.9	&	+	42	:	29	:	39	&		&	faint	&							&							&	0.472	&	$\mathrm{	<hard>	716		}$	\\
267	&	00	:	46	:	09.5	&	+	42	:	15	:	45	&		&	faint	&							&							&	0.221	&	$\mathrm{	<hard>	717		}$	\\
268	&	00	:	46	:	11.4	&	+	41	:	59	:	03	&	pn	&	pl	&	0.3	$_{-	0.3	}^{+	0.6	}$	&	2.3	$_{-	1.3	}^{+	3.2	}$	&	2.58 (1.33)	&	$\mathrm{	<hard>	720		}$	\\
269	&	00	:	46	:	11.7	&	+	42	:	08	:	25	&	pn \& mos	&	pl	&	0.3	$_{-	0.1	}^{+	0.1	}$	&	1.7	$_{-	0.2	}^{+	0.1	}$	&	9.54 (2.04) / 11.1 (1.58)	&	$\mathrm{	<hard>	721		}$	\\
270	&	00	:	46	:	12.1	&	+	42	:	08	:	27	&	pn	&	pl	&	f						&	1.4	$_{-	0.5	}^{+	0.5	}$	&	6.86 (2.77)	&	$\mathrm{	<hard>	721		}$	\\
271	&	00	:	46	:	12.5	&	+	42	:	21	:	51	&	pn \& mos	&	pl	&	0.2	$_{-	0.1	}^{+	0.2	}$	&	1.3	$_{-	0.3	}^{+	0.3	}$	&	2.52 (0.89) / 3.81 (0.88)	&	$\mathrm{	<hard>	723		}$	\\
272	&	00	:	46	:	12.6	&	+	42	:	10	:	26	&		&	faint	&							&							&	0.558	&	$\mathrm{	<hard>	722		}$	\\
273	&	00	:	46	:	13.6	&	+	41	:	50	:	41	&	pn	&	pl	&	f						&	1.8	$_{-	0.7	}^{+	1.0	}$	&	3.38 (1.89)	&	$\mathrm{	<hard>	724		}$	\\
274	&	00	:	46	:	13.6	&	+	42	:	12	:	13	&		&	faint	&							&							&	0.261	&	$\mathrm{			<hard>^1	}$	\\
275	&	00	:	46	:	16.4	&	+	42	:	21	:	28	&	pn \& mos	&	pl	&	f						&	1.6	$_{-	0.3	}^{+	0.4	}$	&	1.24 (0.58) / 1.67 (0.65)	&	$\mathrm{	<hard>	728		}$	\\
276	&	00	:	46	:	18.3	&	+	42	:	25	:	36	&	pn \& mos	&	pl	&	0.18	$_{-	0.12	}^{+	0.11	}$	&	2.1	$_{-	0.4	}^{+	0.7	}$	&	2.57 (0.76) / 2.82 (0.72)	&	$\mathrm{	<hard>	731		}$	\\
277	&	00	:	46	:	18.9	&	+	42	:	15	:	54	&	pn \& mos	&	bb	&	f						&	11	$_{-	8	}^{+	10	}$	&	4.29 (3.94) / 4.22 (3.93)	&	$\mathrm{	<AGN>	732		}$	\\
278	&	00	:	46	:	19.9	&	+	42	:	14	:	41	&	pn \& mos	&	pl	&	0.14	$_{-	0.08	}^{+	0.07	}$	&	1.6	$_{-	0.2	}^{+	0.2	}$	&	5.95 (1.96) / 6.20 (1.14)	&	$\mathrm{	<hard>	735		}$	\\
279	&	00	:	46	:	21.9	&	+	42	:	01	:	46	&		&	faint	&							&							&	0.812	&	$\mathrm{			<hard>^1	}$	\\
280	&	00	:	46	:	24.9	&	+	42	:	04	:	22	&	pn \& mos	&	pl	&	0.33	$_{-	0.03	}^{+	0.03	}$	&	1.98	$_{-	0.08	}^{+	0.09	}$	&	89 (6) / 109 (5)	&	$\mathrm{			<hard>^1	}$	\\
281	&	00	:	46	:	25.3	&	+	42	:	24	:	41	&	pn \& mos	&	pl	&	0.7	$_{-	0.4	}^{+	0.7	}$	&	3.9	$_{-	0.9	}^{+	2.5	}$	&	6.05 (4.51) / 6.56 (4.80)	&	$\mathrm{	<hard>	750		}$	\\
282	&	00	:	46	:	25.5	&	+	42	:	04	:	22	&	pn	&	pl	&	0.22	$_{-	0.05	}^{+	0.08	}$	&	1.74	$_{-	0.14	}^{+	0.22	}$	&	65.4 (6.7)	&	$\mathrm{			<hard>^1	}$	\\
283	&	00	:	46	:	26.9	&	+	42	:	01	:	50	&	pn	&	pl	&	0.12	$_{-	0.02	}^{+	0.03	}$	&	1.41	$_{-	0.08	}^{+	0.08	}$	&	147 (9)	&	$\mathrm{	GlC	752		}$	\\
284	&	00	:	46	:	27.0	&	+	42	:	01	:	51	&	pn \& mos	&	pl	&	0.12	$_{-	0.02	}^{+	0.02	}$	&	1.50	$_{-	0.07	}^{+	0.07	}$	&	146 (10) / 168 (8)	&	$\mathrm{	GlC	752		}$	\\
285	&	00	:	46	:	30.8	&	+	42	:	00	:	38	&		&	faint	&							&							&	1.441	&	$\mathrm{			<hard>^1	}$	\\
286	&	00	:	46	:	32.0	&	+	41	:	51	:	25	&		&	faint	&							&							&	1.500	&	$\mathrm{			<hard>^1	}$	\\
287	&	00	:	46	:	32.4	&	+	42	:	13	:	50	&	mos	&	pl	&	1.5	$_{-	1.4	}^{+	4.6	}$	&	2	$_{-	2	}^{+	8	}$	&	1.49 (1.34)	&	$\mathrm{	<hard>	757		}$	\\
288	&	00	:	46	:	34.7	&	+	42	:	17	:	55	&	pn \& mos	&	pl	&	0.9	$_{-	0.4	}^{+	1.0	}$	&	2.3	$_{-	0.5	}^{+	1.3	}$	&	1.64 (0.86) / 1.88 (0.89)	&	$\mathrm{	<hard>	758		}$	\\
289	&	00	:	46	:	37.4	&	+	42	:	16	:	20	&	pn \& mos	&	br	&	f						&	0.5	$_{-	0.2	}^{+	0.3	}$	&	0.43 (0.35) / 041 (0.35)	&	$\mathrm{	<fg star>	764		}$	\\
290	&	00	:	46	:	40.2	&	+	42	:	25	:	20	&	pn \& mos	&	pl	&	0.25	$_{-	0.12	}^{+	0.11	}$	&	3.6	$_{-	1.2	}^{+	0.7	}$	&	4.81 (2.81) / 4.95 (2.35)	&	$\mathrm{	<fg star>	770		}$	\\
291	&	00	:	46	:	40.7	&	+	41	:	54	:	23	&	pn	&	br	&	3	$_{-	2	}^{+	6	}$	&	0.6	$_{-	0.6	}^{+	7.2	}$	&	18.5 (18.5)	&	$\mathrm{	<AGN>	771		}$	\\
292	&	00	:	46	:	42.5	&	+	42	:	20	:	51	&		&	faint	&							&							&	0.493	&	$\mathrm{	<hard>	775		}$	\\
293	&	00	:	46	:	43.1	&	+	42	:	13	:	37	&	pn	&	pl	&	2	$_{-	2	}^{+	46	}$	&	1.2	$_{-	2.0	}^{+	5.5	}$	&	1.37 (1.29)	&	$\mathrm{	<hard>	776		}$	\\
294	&	00	:	46	:	43.4	&	+	42	:	09	:	48	&	pn \& mos	&	pl	&	f						&	1.1	$_{-	0.3	}^{+	0.4	}$	&	1.89 (1.02) / 1.40 (0.65)	&	$\mathrm{	<hard>	777		}$	\\
295	&	00	:	46	:	45.1	&	+	42	:	27	:	18	&	pn \& mos	&	pl	&	f						&	1.1	$_{-	0.5	}^{+	0.7	}$	&	2.50 (1.61) / 1.18 (0.90)	&	$\mathrm{			<hard>^1	}$	\\
296	&	00	:	46	:	48.0	&	+	42	:	08	:	52	&	pn \& mos	&	pl	&	0.4	$_{-	0.1	}^{+	0.1	}$	&	2.0	$_{-	0.2	}^{+	0.2	}$	&	8.03 (1.41) / 9.77 (1.30)	&	$\mathrm{	<AGN>	784		}$	\\
297	&	00	:	46	:	49.2	&	+	42	:	09	:	30	&	mos	&	pl	&	f						&	4.3	$_{-	1.6	}^{+	2.4	}$	&	2.40 (1.81)	&	$\mathrm{	<fg star>	787		}$	\\
298	&	00	:	46	:	49.4	&	+	42	:	25	:	25	&		&	faint	&							&							&	0.238	&	$\mathrm{	<hard>	788		}$	\\
299	&	00	:	46	:	51.7	&	+	42	:	19	:	49	&	pn \& mos	&	pl	&	0.2	$_{-	0.1	}^{+	0.1	}$	&	1.8	$_{-	0.3	}^{+	0.4	}$	&	2.32 (0.57) / 2.50 (0.51)	&	$\mathrm{	<hard>	789		}$	\\
300	&	00	:	46	:	51.8	&	+	42	:	15	:	5	&	pn \& mos	&	pl	&	f						&	1.8	$_{-	0.3	}^{+	0.4	}$	&	1.54 (0.71) / 1.99 (0.70)	&	$\mathrm{	<hard>	790		}$	\\
301	&	00	:	46	:	52.1	&	+	42	:	17	:	10	&	pn \& mos	&	pl	&	0.3	$_{-	0.2	}^{+	0.4	}$	&	2.2	$_{-	0.8	}^{+	0.7	}$	&	2.4 (1.1) / 1.6 (0.8)	&	$\mathrm{	<hard>	792		}$	\\
302	&	00	:	46	:	53.4	&	+	42	:	19	:	13	&	pn	&	pl	&	f						&	0.9	$_{-	0.6	}^{+	0.6	}$	&	1.06 (0.74)	&	$\mathrm{	<hard>	794		}$	\\
303	&	00	:	46	:	54.4	&	+	42	:	10	:	18	&	pn \& mos	&	pl	&	1.4	$_{-	0.6	}^{+	0.9	}$	&	2.2	$_{-	0.5	}^{+	0.9	}$	&	4.33 (1.52) / 4.80 (1.91)	&	$\mathrm{	<hard>	795		}$	\\
304	&	00	:	46	:	55.2	&	+	42	:	20	:	48	&	pn \& mos	&	pl	&	0.15	$_{-	0.01	}^{+	0.01	}$	&	2.25	$_{-	0.06	}^{+	0.07	}$	&	43 (2) / 53 (2)	&	$\mathrm{	<hard>	796		}$	\\
305	&	00	:	46	:	58.4	&	+	42	:	24	:	13	&	pn	&	pl	&	f						&	1.1	$_{-	0.7	}^{+	0.8	}$	&	2.03 (1.40)	&	$\mathrm{	<fg star>	800		}$	\\
306	&	00	:	46	:	59.6	&	+	42	:	18	:	06	&		&	faint	&							&							&	0.282	&	$\mathrm{			<hard>^1	}$	\\
307	&	00	:	47	:	00.4	&	+	42	:	21	:	55	&	pn	&	bb	&	f						&	0.15	$_{-	0.02	}^{+	0.03	}$	&	0.86 (0.53)	&	$\mathrm{	<fg star>	802		}$	\\
308	&	00	:	47	:	01.9	&	+	42	:	22	:	46	&	pn \& mos	&	pl	&	f						&	1.7	$_{-	0.4	}^{+	0.5	}$	&	1.09 (0.58) / 0.86 (0.39)	&	$\mathrm{	<hard>	804		}$	\\
309	&	00	:	47	:	02.6	&	+	42	:	18	:	35	&	pn \& mos	&	pl	&	0.2	$_{-	0.2	}^{+	0.2	}$	&	2.5	$_{-	1.0	}^{+	0.6	}$	&	1.26 (0.38) / 1.28 (0.39)	&	$\mathrm{	<hard>	805		}$	\\
310	&	00	:	47	:	03.6	&	+	42	:	04	:	48	&	pn \& mos	&	pl	&	0.2	$_{-	0.1	}^{+	0.2	}$	&	1.6	$_{-	0.3	}^{+	0.4	}$	&	7.90 (2.04) / 6.91 (1.45)	&	$\mathrm{	<hard>	808		}$	\\
311	&	00	:	47	:	04.3	&	+	42	:	16	:	47	&		&	faint	&							&							&	0.533	&	$\mathrm{	<hard>	812		}$	\\
312	&	00	:	47	:	06.3	&	+	42	:	22	:	09	&	pn \& mos	&	pl	&	0.41	$_{-	0.12	}^{+	0.19	}$	&	1.8	$_{-	0.2	}^{+	0.3	}$	&	5.02 (1.03) / 6.27 (0.95)	&	$\mathrm{	<hard>	817		}$	\\
313	&	00	:	47	:	07.6	&	+	42	:	18	:	10	&	pn \& mos	&	pl	&	0.6	$_{-	0.2	}^{+	0.5	}$	&	1.9	$_{-	0.3	}^{+	0.6	}$	&	2.65 (0.81) / 2.77 (0.78)	&	$\mathrm{	<hard>	820		}$	\\
314	&	00	:	47	:	08.6	&	+	42	:	24	:	04	&	pn \& mos	&	pl	&	0.39	$_{-	0.15	}^{+	0.28	}$	&	2.0	$_{-	0.4	}^{+	0.2	}$	&	4.21 (1.21) / 3.03 (0.88)	&	$\mathrm{	<hard>	821		}$	\\
315	&	00	:	47	:	09.0	&	+	42	:	10	:	09	&	mos	&	bb	&	f						&	2	$_{-	1	}^{+	198	}$	&	11.2 (10.2)	&	$\mathrm{	<hard>	822		}$	\\
316	&	00	:	47	:	10.5	&	+	42	:	18	:	45	&	mos	&	pl	&	f						&	1.7	$_{-	0.6	}^{+	0.8	}$	&	1.10 (0.64)	&	$\mathrm{	<hard>	824		}$	\\
317	&	00	:	47	:	10.8	&	+	42	:	16	:	13	&	pn	&	pl	&	f						&	2.3	$_{-	1.4	}^{+	1.9	}$	&	1.09 (0.98)	&	$\mathrm{	<hard>	825		}$	\\
318	&	00	:	47	:	11.3	&	+	42	:	22	:	22	&	pn \& mos	&	pl	&	3.4	$_{-	1.5	}^{+	2.6	}$	&	3.0	$_{-	0.7	}^{+	1.4	}$	&	5.85 (4.03) / 8.87 (6.45)	&	$\mathrm{	<hard>	826		}$	\\
319	&	00	:	47	:	13.2	&	+	42	:	20	:	44	&	pn \& mos	&	pl	&	0.14	$_{-	0.06	}^{+	0.06	}$	&	2.1	$_{-	0.2	}^{+	0.2	}$	&	6.97 (1.10) / 8.27 (0.89)	&	$\mathrm{	<hard>	827		}$	\\
320	&	00	:	47	:	21.0	&	+	42	:	05	:	47	&	pn \& mos	&	br	&	0.48	$_{-	0.42	}^{+	0.17	}$	&	0.18	$_{-	0.09	}^{+	0.25	}$	&	9.07 (9.07) / 5.65 (5.65)	&	$\mathrm{	<fg star>	835		}$	\\
321	&	00	:	47	:	24.0	&	+	42	:	08	:	43	&		&	faint	&							&							&	0.402	&	$\mathrm{	<hard>	837		}$	\\
322	&	00	:	47	:	25.2	&	+	42	:	21	:	16	&	pn \& mos	&	pl	&	0.19	$_{-	0.09	}^{+	0.07	}$	&	2.2	$_{-	0.3	}^{+	0.2	}$	&	3.15 (0.74) / 3.95 (0.77)	&	$\mathrm{	<hard>	839		}$	\\
323	&	00	:	47	:	26.0	&	+	42	:	21	:	57	&	pn \& mos	&	br	&	0.96	$_{-	0.05	}^{+	0.05	}$	&	0.12	$_{-	0.02	}^{+	0.03	}$	&	538 (524) / 605 (593)	&	$\mathrm{	<fg star>	840		}$	\\
324	&	00	:	47	:	27.4	&	+	42	:	13	:	46	&	pn \& mos	&	br	&	2.7	$_{-	1.5	}^{+	3.3	}$	&	1.4	$_{-	1.4	}^{+	2.6	}$	&	2.18 (2.18) / 2.57 (2.57)	&	$\mathrm{	<hard>	842		}$	\\
325	&	00	:	47	:	30.5	&	+	42	:	12	:	01	&		&	faint	&							&							&	0.611	&	$\mathrm{			<hard>^1	}$	\\
326	&	00	:	47	:	35.9	&	+	42	:	08	:	34	&	pn \& mos	&	pl	&	0.12	$_{-	0.12	}^{+	0.31	}$	&	1.3	$_{-	0.6	}^{+	0.9	}$	&	3.30 (1.72) / 3.40 (1.43)	&	$\mathrm{	<hard>	845		}$	\\
327	&	00	:	47	:	38.4	&	+	42	:	20	:	21	&	mos	&	pl	&	0.11	$_{-	0.11	}^{+	0.25	}$	&	1.4	$_{-	0.5	}^{+	0.7	}$	&	4.78 (1.42)	&	$\mathrm{	<AGN>	846		}$	\\
328	&	00	:	47	:	42.5	&	+	42	:	22	:	27	&	mos	&	pl	&	f						&	1.9	$_{-	0.4	}^{+	0.5	}$	&	3.92 (1.43)	&	$\mathrm{	<hard>	848		}$	\\
329	&	00	:	47	:	42.6	&	+	42	:	11	:	38	&	pn \& mos	&	pl	&	0.2	$_{-	0.2	}^{+	0.3	}$	&	1.8	$_{-	0.6	}^{+	0.6	}$	&	1.93 (0.95) / 2.09 (0.72)	&	$\mathrm{	<hard>	849		}$	\\
330	&	00	:	47	:	42.7	&	+	42	:	10	:	15	&		&	faint	&							&							&	0.618	&	$\mathrm{	<hard>	850		}$	\\
331	&	00	:	47	:	43.7	&	+	42	:	12	:	19	&		&	faint	&							&							&	0.747	&	$\mathrm{	<hard>	851		}$	\\
332	&	00	:	47	:	44.4	&	+	42	:	10	:	58	&	pn \& mos	&	pl	&	0.23	$_{-	0.16	}^{+	0.22	}$	&	2.0	$_{-	0.5	}^{+	0.4	}$	&	2.52 (0.98) / 3.93 (1.14)	&	$\mathrm{	<hard>	852		}$	\\
333	&	00	:	47	:	46.5	&	+	42	:	18	:	49	&		&	faint	&							&							&	0.821	&	$\mathrm{			<hard>^1	}$	\\
334	&	00	:	47	:	46.9	&	+	42	:	14	:	22	&	mos	&	pl	&	0.7	$_{-	0.5	}^{+	1.1	}$	&	2.7	$_{-	1.6	}^{+	1.9	}$	&	3.72 (2.71)	&	$\mathrm{	<hard>	853		}$	\\
335	&	00	:	47	:	48.1	&	+	42	:	19	:	33	&	mos	&	pl	&	0.21	$_{-	0.08	}^{+	0.09	}$	&	1.6	$_{-	0.2	}^{+	0.2	}$	&	60.8 (5.30)	&	$\mathrm{	<AGN>	855		}$	\\

  \end{longtable}

\end{landscape}
\end{center}
\end{appendix}

\end{document}